\documentclass[draft]{agujournal2019}
\usepackage{url} %this package should fix any errors with URLs in refs.
\usepackage{lineno}
\usepackage[inline]{trackchanges} %for better track changes. finalnew option will compile document with changes incorporated.
\usepackage{soul}
\usepackage{color}

\draftfalse

\journalname{JGR: Planets}

\begin{document}

%% ------------------------------------------------------------------------ %%
%  Title
%
% (A title should be specific, informative, and brief. Use
% abbreviations only if they are defined in the abstract. Titles that
% start with general keywords then specific terms are optimized in
% searches)
%
%% ------------------------------------------------------------------------ %%

% Example: \title{This is a test title}

\title{The Volcanic and Radial Expansion/Contraction History of the Moon Simulated by Numerical Models of Magmatism in the Convective Mantle}

%% ------------------------------------------------------------------------ %%
%
%  AUTHORS AND AFFILIATIONS
%
%% ------------------------------------------------------------------------ %%

% Authors are individuals who have significantly contributed to the
% research and preparation of the article. Group authors are allowed, if
% each author in the group is separately identified in an appendix.)

% List authors by first name or initial followed by last name and
% separated by commas. Use \affil{} to number affiliations, and
% \thanks{} for author notes.
% Additional author notes should be indicated with \thanks{} (for
% example, for current addresses).

% Example: \authors{A. B. Author\affil{1}\thanks{Current address, Antartica}, B. C. Author\affil{2,3}, and D. E.
% Author\affil{3,4}\thanks{Also funded by Monsanto.}}

\authors{Ken'yo U\affil{1}\thanks{“U” is the family name in the author Ken’yo U.}, Masanori Kameyama\affil{2}, and Masaki Ogawa\affil{1}}

 \affiliation{1}{Department of Earth Sciences and Astronomy, The University of Tokyo, Komaba, Meguro, Japan}
 \affiliation{2}{Geodynamics Research Center, Ehime University, Matsuyama, Japan}
% \affiliation{3}{Third Affiliation}
% \affiliation{4}{Fourth Affiliation}

%\affiliation{=number=}{=Affiliation Address=}
%(repeat as many times as is necessary)

%% Corresponding Author:
% Corresponding author mailing address and e-mail address:

% (include name and email addresses of the corresponding author.  More
% than one corresponding author is allowed in this LaTeX file and for
% publication; but only one corresponding author is allowed in our
% editorial system.)

% Example: \correspondingauthor{First and Last Name}{email@address.edu}

\correspondingauthor{Ken'yo U}{u-kenyo0822@g.ecc.u-tokyo.ac.jp}

%% Keypoints, final entry on title page.

%  List up to three key points (at least one is required)
%  Key Points summarize the main points and conclusions of the article
%  Each must be 140 characters or fewer with no special characters or punctuation and must be complete sentences

% Example:
% \begin{keypoints}
% \item	List up to three key points (at least one is required)
% \item	Key Points summarize the main points and conclusions of the article
% \item	Each must be 140 characters or fewer with no special characters or punctuation and must be complete sentences
% \end{keypoints}

\begin{keypoints}
\item We numerically simulated magmatism in the convecting mantle of the Moon to understand its evolution
\item Magma is generated in the deep mantle and ascends to the surface as partially molten fingers and plumes driven by melt-buoyancy
\item Extension of partially molten regions by the fingers and plumes cause the observed expansion and active volcanism of the early Moon
\end{keypoints}

%% ------------------------------------------------------------------------ %%
%
%  ABSTRACT and PLAIN LANGUAGE SUMMARY
%
% A good Abstract will begin with a short description of the problem
% being addressed, briefly describe the new data or analyses, then
% briefly states the main conclusion(s) and how they are supported and
% uncertainties.

% The Plain Language Summary should be written for a broad audience,
% including journalists and the science-interested public, that will not have 
% a background in your field.
%
% A Plain Language Summary is required in GRL, JGR: Planets, JGR: Biogeosciences,
% JGR: Oceans, G-Cubed, Reviews of Geophysics, and JAMES.
% see http://sharingscience.agu.org/creating-plain-language-summary/)
%
%% ------------------------------------------------------------------------ %%

%% \begin{abstract} starts the second page

\begin{abstract}
[To understand the evolution of the Moon, we numerically modeled mantle convection and magmatism in a two-dimensional polar rectangular mantle. Magmatism occurs as an upward permeable flow of magma generated by decompression melting through the convecting matrix. The mantle is assumed to be initially enriched in heat-producing elements (HPEs) and compositionally dense ilmenite-bearing cumulates (IBC) at its base. 
Here, we newly show that magma generation and migration play a crucial role in the calculated volcanic and radial expansion/contraction history. Magma is generated in the deep mantle by internal heating for the first several hundred million years. A large volume of the generated magma ascends to the surface as partially molten fingers and plumes driven by melt-buoyancy to cause a volcanic activity and radial expansion of the planet with the peak at 3.5-4 Gyr ago. Eventually, however, the planet begins to radially contract when the mantle solidifies by cooling from the surface boundary. As the mantle is cooled, the activity of partially molten plumes declines but continues for billions of years after the peak because some basal materials enriched in the dense IBC components hold HPEs. The calculated volcanic and radial expansion/contraction history is consistent with the observed history of the Moon. Our simulations suggest a substantial fraction of the mantle was solid, and there was a basal layer enriched in HPEs and the IBC components at the beginning of the history of the Moon.]
\end{abstract}

\section*{Plain Language Summary}
[We developed a numerical model of magmatism in the convecting mantle to understand the volcanic and radial expansion/contraction history of the Moon. In the early period of the calculated history, magma is generated in the deep mantle and ascends to the surface as partially molten fingers and plumes driven by melt-buoyancy to cause volcanic activity. The extension of partially molten regions by magma ascent causes radial expansion of the planet. In its later period, however, the planet contracts with time because partially molten regions solidify as they are cooled from the surface boundary. The activity of partially molten plumes declines but continues for billions of years because some materials that host heat-producing elements are enriched in a compositionally dense component and remain in the deep mantle. The calculated history of radius change and volcanism is consistent with the observed lunar history. Our simulations suggest that a substantial fraction of the mantle was solid, and a dense layer enriched in heat-producing elements developed at the base of the mantle at the beginning of the history of the Moon.]

%% ------------------------------------------------------------------------ %%
%
%  TEXT
%
%% ------------------------------------------------------------------------ %%

%%% Suggested section heads:
% \section{Introduction}
%
% The main text should start with an introduction. Except for short
% manuscripts (such as comments and replies), the text should be divided
% into sections, each with its own heading.

% Headings should be sentence fragments and do not begin with a
% lowercase letter or number. Examples of good headings are:

% \section{Materials and Methods}
% Here is text on Materials and Methods.
%
% \subsection{A descriptive heading about methods}
% More about Methods.
%
% \section{Data} (Or section title might be a descriptive heading about data)
%
% \section{Results} (Or section title might be a descriptive heading about the
% results)
%
% \section{Conclusions}

\section{Introduction}

 % \begin{figure}
 %\noindent\includegraphics[width=\textwidth]{sanmple.png}
%\caption{caption}
%\label{pngfiguresample}
%\end{figure}
%Text here ===>>>

\noindent Understanding the mantle evolution of the Moon that is behind its observed history of volcanic activity and radius change has been a long-standing issue in studies of the interiors of terrestrial planets \cite<e.g.,>{s&c1976,kirk&s,Shearer,Breuer}. The Moon expanded globally by 0.5-5 km in its earlier history until around 3.8 Gyr ago as revealed by the gravity gradiometry data \cite{hana2013,hana2014,sawada,liang&Andrews}, and it then globally contracted until today, as suggested from observations of tectonic features on the Moon \cite{Yue,Frueh}; some observations of fault scarps (thrust faults) suggest that the contraction for the past 100 Myr is around 1 km or less \cite<e.g.,>{watters,watters2015,Klimczak,clark,bogert,matsu}.
The period when its radius reached the maximum coincides with that when the mare volcanism was active: mare volcanism became more active with time for the first several hundred million years of the lunar history, peaked at 3.5-3.8 Gyr ago, and then declined but continued until around 1.5 Gyr ago \cite<e.g.,>{hiesinger2000,hiesinger2003,morota2011b,whitten&head}. 
To clarify the mantle evolution that has caused the observed features of the lunar history, we developed a two-dimensional (2-D) polar rectangular model of the lunar mantle evolution where mantle convection and magmatism that transport heat, mass, and incompatible heat-producing elements (HPEs) are considered.

Various numerical models of mantle evolution have been advanced to account for the observed radius change, in particular, the early expansion of the Moon \cite<e.g.,>{Solomon1986,Shearer}. Classical spherically symmetric one-dimensional (1-D) models where the radius change occurs only thermally show that the early expansion is reproduced in the models when the temperature in the deep mantle is initially 1100 K or less; these models suggest that subsequent temperature rise of more than 700 K in the deep mantle due to internal heating caused the observed early expansion of the Moon \cite{s&c1976}. 
In the model of \citeA{kirk&s} that also takes volume change from compositional differentiation of the mantle into account, the early expansion occurs even when the initial temperature of the deep mantle is as high as 1200 K.
 Giant impact hypotheses for the origin of the Moon \cite<e.g.,>{Stevenson,p&s,canup,Cuk&Stewart,Rufu,lock}, however, suggest a much higher initial temperature for the Moon. The mantle is expected to have been mostly molten to form the magma ocean immediately after the impact \cite<e.g.,>{Newsom&Taylor,hosono}. Even at the end of the mantle overturn, which is expected to have occurred after solidification of the magma ocean, the temperature of the deep mantle is still suggested to have been as high as about 1800-1900 K \cite<e.g.,>{hess&P,hess&P2001,alley,bouk,li2019}.
When such a high initial temperature is assumed, the early expansion of the Moon is difficult to reproduce in the classical thermal history models \cite<e.g.,>{s&c1976,Solomon1986}. The early expansion consistent with the observed history does not occur in three-dimensional (3-D) spherical models where mantle convection is also considered, too. In the models of \citeA{zhang2013a,zhang2013b}, the early expansion does occur owing to internal heating, but the amplitude of the expansion is much smaller than the observed value \cite<e.g.,>{hana2013}, and the period when the expansion occurs is longer than 1 Gyr.
 In some models where the uppermost mantle is locally more enriched in HPEs on the nearside as observed for the Procellarum KREEP terrane, or PKT \cite{PKT,Laneuville2013}, the expansion occurs only on the nearside and is too large to account for the observed expansion \cite<e.g.,>{liang&Andrews}.
 To understand the observed early expansion of the Moon, \citeA{u2022} constructed a 1-D spherically symmetric model where volume change of the mantle by melting is considered in addition to thermal expansion. The mantle expands by a few kilometers for the first several hundred million years of the calculated history when partially molten regions extend in the mantle by internal heating, suggesting that melting of the mantle played an important role in the early expansion of the Moon. Yet, mantle convection is not considered in this model.

Various numerical models have been advanced to account for the long-lasting volcanism of the Moon, too \cite<e.g.,>{Shearer&Papike,Breuer}. 
Some earlier models show that partially molten regions persist for billions of years in the upper mantle when the surface is covered with the crust enriched in HPEs or the blanketing regolith layer \cite{k&s,Spohn,zie}.
In models where the uppermost mantle in the nearside is more enriched in HPEs than that in the farside, the mantle has been partially molten for more than 3 billion years \cite{Laneuville2013,Laneuville2014,Laneuville2018}. In these models, however, the distribution of HPEs is spatially fixed or transported only by mantle convection, and extraction of HPEs from the mantle to the crust by magmatism, which is known to reduce the activity of magmatism \cite<e.g.,>{cassen1973,cassen1979,ogawa2014}, is not considered.
In a model of \citeA{ogawa2018B} where extraction of HPEs from the mantle by migrating magma is considered, partially molten regions in the mantle are all solidified, and magmatism stops within the first 2 Gyr of the calculated history, too early to account for the lunar mare vocanism \cite<e.g.,>{hiesinger2003}.
It is also important to consider not only HPE-extraction by migrating magma, but also the effects of structural evolution of the mantle in studies of the lunar volcanism \cite<e.g.,>{hess&P,zhong,stegman,zhang2013a,zhang2022}. In the early period of the Moon, a compositionally dense layer, which is enriched in HPEs and ilmenite-bearing cumulates (IBC) components, is suggested to have developed at the base of the mantle by crystal fractionation in the magma ocean and subsequent mantle overturn \cite<e.g.,>{R&K,Elkins-Tanton,moriarty}. Earlier mantle convection models suggest that the basal layer becomes thermally buoyant owing to internal heating and eventually migrates upward to the surface as hot plumes to cause mare volcanism after around 4 Gyr ago \cite<e.g.,>{stegman}. 
 Whether or not the basal layer rises, however, depends on the compositional density contrast between the layer and the overlying mantle. These models assume conditions where the density contrast is low enough to allow the basal layer to rise by thermal convection (see Figure 1 in \citeA{Bars}). The density contrast after mantle overturn is unclear and is influenced by the initial condition of overturn models \cite<e.g.,>{li2019,yu,zhang2022}; further studies with various density content are necessary to fully understand to effects of compositional mantle structure on volcanic history.

%Some laboratory deformation experiments imply that materials in the overturned layer do not ascend by mantle convection alone in most cases of the Moon \cite{dygert,tokle}.

To understand the mantle evolution in the Moon that is constrained by its volcanic and radial expansion/contraction history, we extend the 1-D spherically symmetric model we developed earlier \cite{u2022} to a 2-D polar rectangular model. In our previous model, we considered magma generation by internal heating and magma migration that transports heat, HPE, and mass as well as volume change of the mantle by melting. 
In addition to these effects, here in this study, we also include the effects of mantle convection. Heat and mass transport by mantle convection plays an important role in mantle evolution \cite<e.g.,>{Spohn,zhang2017}. Upwelling flows of mantle convection generate magma by decompression melting, which also affects mantle evolution of the planet \cite<e.g.,>{ogawa2020}. We also take the effects of compositional layering at the base of the mantle formed by the magma ocean and mantle overturn on the subsequent mantle evolution into account. This model is an extension of the one presented in \citeA{ogawa2014,ogawa2018B} in that we considered the volume change of mantle by melting, more systematically studied the effects of the initial condition, and calculated in a 2-D polar rectangle rather than rectangle.

% The location of the PKT has around 8 km thermal contraction between 4.0 and 3.0 Gyr ago as a result of the declining radiogenic heat production \cite{hana2014}.
 
\section{Model description}
\noindent A finite difference numerical code calculates the energy, mass, and momentum equations for mantle magmatism and mantle convection in a two-dimensional polar rectangular $R=\left[ \left( r, \theta \right) | \; 385 \; \mathrm{km} \leq r \leq 1735 \; \mathrm{km} , \; 0 \leq \theta \leq \pi \right]$ under the Boussinesq approximation, where the inner and the outer radii correspond to the core and planetary radii, respectively, of the Moon \cite <e.g.,>{weber2011,yan2015,viswanathan}. The mantle contains incompatible heat-producing elements (HPEs) that decay with time. Mantle convection occurs as a convection of a Newtonian fluid whose viscosity strongly depends on temperature and is driven by thermal, compositional, and melt-buoyancy. The convecting materials are a binary eutectic system between olivine-rich materials and ilmenite-bearing cumulates (IBC). Magmatism occurs as generation of basaltic magma enriched in IBC materials and HPEs by decompression melting, internal heating, and upward permeable flow of the generated magma through the matrix \cite{VanderKaaden,Mc1984}; the permeable flow is driven by the melt-buoyancy. 

%\color{blue}
% This expectation is based on the inference that almost melt compositions of the lunar picritic glasses should have been able to rise to the uppermost mantle obtained from sink-float experiments \cite{VanderKaaden}.

The crust of 35 km in thickness is placed on the top of the mantle \cite{wieczorek2013}. 
The thermal diffusivity of the crust is about half that of the mantle, and hence the crust serves as a blanketing layer \cite{zie}. The temperature is fixed at $T_{\mathrm{sur}} = 270$ K on the surface boundary, while the core is modeled as a heat bath of uniform temperature; the vertical sidewalls are insulating. All of the boundaries are impermeable for both magma and matrix and are shear stress-free. In the initial condition, we assume that the deep mantle is more enriched in HPEs and IBC component than the shallower mantle as earlier models of mantle differentiation by the magma ocean and subsequent mantle overturn suggest \cite <e.g.,>{R&K,hess&P,moriarty}. Several previous studies, however, suggest that a portion of the IBC materials enriched in HPEs persist just beneath the crust even after mantle overturn \cite<e.g.,>{yu,zhao2019,Schwinger}. Instead of explicitly simulating these remains of HPE-enriched materials, we assume that the crust is uniformly enriched in HPEs.

The numerical mesh employed for the calculation contains 128 (radial direction) times 256 (lateral direction) mesh points.

\begin{table}
 \caption{The meanings of the symbols and their values.}
 \label{symbol}
 \centering
 \begin{tabular}{l l l}
 \hline
  Symbol  & Meaning  & Value \\
 \hline
   $T_{\mathrm{sur}}$  & Surface temperature  & $270$ K \\
   $T_0$  & Solidus at the surface  & $1360$ K   \\
   $r_p$  & Radius of the surface  & $1735$ km \\
   %   $r_p$  & Radius of the surface  & $1731$ km \\
   $r_c$  & Radius of the core-mantle boundary (CMB)  & $385$ km   \\
   %   $r_c$  & Radius of the core-mantle boundary  & $381$ km   \\
   $r_{crst}$  & Radius of the Moho  & $1695$ km   \\
   %   $r_{crst}$  & Radius of the Moho  & $1694$ km   \\
   $\rho_0$  & Reference density  & $3300$ kg $\mathrm{m^{-3}}$   \\
   $\kappa$  & Thermal diffusivity of the mantle  & $6 \times 10^{-7}$ $\mathrm{m^2}$ $\mathrm{s^{-1}}$   \\
    $\kappa_{edd}$  & Eddy diffusivity in largely molten region  & $100 \kappa$ at $\phi > 0.4$  \\
   ${g_{\mathrm{sur}}}$   & Gravitational acceleration at the surface  & $1.62$ m $\mathrm{s^{-2}}$    \\
      ${g_{\mathrm{c}}}$   & Gravitational acceleration at the CMB  & $0.55$ m $\mathrm{s^{-2}}$    \\
   $\Delta h$   & Latent heat of melting  & $657$ kJ $\mathrm{Kg^{-1}}$  \\
   $C_p$   & Specific heat  & $1240$ J $\mathrm{K^{-1}}$ $\mathrm{kg^{-1}}$    \\
   $\eta_{0}$  & Reference viscosity  & $10^{20 } - 10^{22}$ Pa s   \\
   $\eta_{\mathrm{melt}}$  & Melt viscosity  & $1 - 20 $ Pa s   \\
       $E_T$  & Sensitivity of viscosity to temperature  & $11.3 \times 10^{-3}$ $\mathrm{K}^{-1}$   \\
    $\phi_0$  & Reference melt-content  & $0.05$   \\
     $k_{\mathrm{\phi_{\mathrm{0}}}}$  & Reference permeability  & $8.6 \times 10^{-15}$ $\mathrm{m^2}$   \\
     $\alpha_{\mathrm{m}}$  & Thermal expansivity in the mantle  & $3 \times 10^{-5}$ $\mathrm{K^{-1}}$   \\
    $\alpha_{\mathrm{c}}$  & Thermal expansivity in the core  & $9 \times 10^{-5}$ $\mathrm{K^{-1}}$   \\
 \hline
% \multicolumn{2}{l}{$^{a}$Footnote text here.}
 \end{tabular}
 \end{table}

\subsection{The properties of materials}
\noindent The convecting material is a binary eutectic system. The composition is written as $A_{\xi} B_{1-{\xi}}$; the mantle materials are modeled as a mixture of olivine-rich materials $A$ with a density of $3300$ kg $\mathrm{m^{-3}}$ \cite{Elkins-Tanton} and the IBC component $B$ with a density of $3745$ kg $\mathrm{m^{-3}}$ \cite{snyder,Shearer,Elkins-Tanton,rapp}. The eutectic composition is $A_{0.1} B_{0.9}$ which corresponds to the composition of the basaltic composition enriched in the IBC component (see Appendix for the detail of the thermodynamic formulation).

The density $\rho$ is written as
    \begin{linenomath*}
    \begin{equation}
     \rho = \left( 1-\phi\right) \rho_\mathrm{s} + \phi \rho_\mathrm{l}\,.
   \end{equation}
   \end{linenomath*}
   \noindent where $\rho_\mathrm{s}$ is the density of solid-phase, $\rho_\mathrm{l}$ the density of melt-phase, and $\phi$ the melt-content. The densities depend on the temperature $T$ and the content of the end-member $A$ in the solid-phase $\xi_\mathrm{s}$ and that in the melt-phase $\xi_\mathrm{l}$ as
    \begin{linenomath*}
    \begin{equation}
      \rho_{\mathrm{s}} = \rho_{\mathrm{0}}\left[1-\alpha \left( T-T_{\mathrm{sur}}\right)+\beta \left(1-\xi_{\mathrm{s}} \right)\right]  \,,
   \end{equation}
   and
   \end{linenomath*}
       \begin{linenomath*}
    \begin{equation}
      \rho_{\mathrm{l}} = \rho_{\mathrm{0}}\left\{ 1-\alpha \left( T-T_{\mathrm{sur}}\right)+\beta \left(1-\xi_{\mathrm{l}} \right) - \frac{\Delta v_{\mathrm{l}}}{v_{\mathrm{0}}}\left[ 1+
      \beta\left(1-\xi_{\mathrm{l}}\right)\right]\right\} \,,
   \end{equation}
   \end{linenomath*}
   \noindent where $\rho_0$ is the reference density; $\alpha$ the thermal expansivity; $\beta = 0.135$ a constant estimated from the density of olivine-rich end-member A and that of the IBC end-member B. Temperature is calculated by the `reduced’ enthalpy $h$ which is defined as
   \begin{equation}
   \label{enthalpy}
      h = C{_{p}} T +
      \phi \Delta h \left( 1+ G\right)\,,
   \end{equation}
   where $C{_{p}}$ is the specific heat; $\Delta h$ the latent heat of melting; $G$ the function which depends on ${\Delta v_{\mathrm{l}}}/{v_{\mathrm{0}}}$ (see Eq. \ref{g_eq} below). 
    ${\Delta v_{\mathrm{l}}}/{v_{\mathrm{0}}}$ expresses the amount of density reduction by melting as
    \begin{linenomath*}
   \begin{equation}
      \frac{\Delta v_{\mathrm{l}}}{v_{\mathrm{0}}} =
      \frac{1}{\rho_0} \left[ \Delta \rho_\infty+\frac{\Delta \rho_{\mathrm{zero}} -\Delta \rho_\infty}{\left( P/\lambda +1\right) ^2} \right]
      %D_{\mathrm{0}} +  {D_{\mathrm{1}}}/{\left[ 1+( r_p -r)/ \iota \right]^2}
                 \,.
   \end{equation}
     \end{linenomath*}

     %here 1014%
\noindent Here, $\Delta \rho_\infty/\rho_0=0.005$ is the dimensionless density difference between solid- and liquid-phases when pressure is infinity, and $\Delta \rho_{\mathrm{zero}}/\rho_0=0.22$
%$\Delta \rho_n/\rho_0=0.250$
that at zero-pressure. $P$ is the pressure which difined as $P = - \int_{r_p}^r \rho_0 {g} dr$; the gravitational acceleration $g$ depends on the depth as $g= g_{\mathrm{sur}}- \left (g_{\mathrm{sur}}- g_{\mathrm{c}}\right) \frac{r_p -r}{r_p - r_c}$ \cite{garciaA,garciaB}. We assumed the value of $\lambda = 16.42 \; \mathrm{G Pa}$ so that the solidus temperature is calculated from the Clausius-Clapeyron relationship (see Eq. \ref{mt_eq}, \ref{g_eq} below) becomes close to that in the lunar mantle \cite{katz,garciaA,garciaB}.
%where $r$ is the radial coordinate, and the values of $D_{0}$, $D_{1}$, $\iota$ which determine the solidus curve and the value of sensitivity of density to composition $\beta = 0.135$  are chosen as same as \citeA{u2022}.

Magma is generated by decompression melting and internal heating. The generated magma migrates upward as a permeable flow through the coexisting matrix driven by its buoyancy. Migrating magma transports heat, basaltic materials, and HPEs. The difference between the velocity of magma $\mathbf{u}$ and that of matrix $\mathbf{U}$ is proportional to the density difference between them as
     \begin{linenomath*}
    \begin{equation}
     \label{relative}
      \mathbf{u}-\mathbf{U} =
       \frac{k_{\mathrm{\phi}}}
        {\phi \eta_{\mathrm{melt}}} {g} \left(\rho_{s} - \rho_{l} \right) \mathbf{e_r}
       \,,
   \end{equation}
     \end{linenomath*}
%  Numbered lines in equations:
%  To add line numbers to lines in equations,
%  \begin{linenomath*}
%  \begin{equation}
%  \end{equation}
%  \end{linenomath*}
where $k_{\mathrm{\phi}}$ is the permeability that depends on the melt-content $\phi$ as
     \begin{linenomath*}
    \begin{equation}
    \label{k_phi}
      k_{\mathrm{\phi}}=k_{\mathrm{\phi_{\mathrm{0}}}}
        \left( \frac{ \phi}
        {\phi_{\mathrm{0}}} \right)^3
       \,
   \end{equation}
     \end{linenomath*}
 \cite{Mc1984} and $\mathbf{e_r}$ is the unit vector in the radial direction. We assumed that $\phi_0 = 0.05$ and truncated $\phi$ at 0.4 for a numerical reason. The assumed range of $k_{\mathrm{\phi_{\mathrm{0}}}}$ given in Table \ref{symbol} is based on the earlier works of  \citeA{Mc1984} and \citeA{miller}. In the top-most 150 km of the polar rectangular, we inserted $\phi' = \mathrm{max} \left(\phi, \phi_{e} \right)$ into the $\phi$ in Eq. \ref{k_phi} to mimic magma migration that occurs by crack propagation in the crust and the uppermost mantle of the Moon \cite{HW, WH}
      \begin{linenomath*}
    \begin{equation}
      k_{\mathrm{\phi}}=k_{\mathrm{\phi_{\mathrm{0}}}}
         \frac{ \left[\mathrm{max} \left(\phi, \phi_{e} \right) \right]^3}
        {\phi_{\mathrm{0}}^3}
       \,,
   \end{equation}
     \end{linenomath*}
where $\phi_{e} = 0.035$ \cite{Kameyama}. 

The solidus temperature ${T_{\mathrm{solidus}}}$ depends on the pressure as
       \begin{linenomath*}
    \begin{equation}
     \label{mt_eq}
     T_{\mathrm{solidus}}= T_{\mathrm{0}} \left(  1+G \right)
       \,,
   \end{equation}
where 
     \end{linenomath*}
           \begin{linenomath*}
    \begin{equation}
    \label{g_eq}
     G= \frac{1}{\rho_{\mathrm{0}} \Delta h} 
     \int_0^P \frac{\Delta v_{\mathrm{l}}}{v_{\mathrm{0}}}\,dP\,.
   \end{equation}
     \end{linenomath*}
     Here, $T_{\mathrm{0}}$ is the solidus temperature at the surface \cite{katz}.
%%

%Kameyama has inserted a linebreak in order to start a new paragraph.
The viscosity of the mantle $\eta$ depends on temperature as
     \begin{linenomath*}
    \begin{equation}
    \label{T-dep}
     \eta = \eta_{0} \exp{\left[ E_{T} \left( T_\mathrm{ref} - T \right)\right]} \,,
   \end{equation}
   \end{linenomath*}
  \noindent where $\eta_{0}$ is the reference viscosity; $T_\mathrm{ref} = 1575  \;  \mathrm{K}$ is the reference temperature. $E_{T}$ is the sensitivity of viscosity to temperature; we adopt the default value of $E_{T}$ is $11.3 \times 10^{-3}$ $\mathrm{K}^{-1}$ ($E_T^*=6$, see Section  \ref{The parameter values} below) which implies that the viscosity decreases by a factor of 3 as temperature increases by 100 K. This range of viscosity is appropriate for Newtonian rheology to mimic thermal convection in mantle materials \cite{Dumoulin}. The value of $E_{T}$ also prompts the ratio of viscosity at the surface boundary to that at the Moho discontinuity is a factor 5, which is a same value as that in some earlier studies obtained by Arrhenius temperature-dependence \cite<e.g.,>{zhang2013a,zhang2017}.
  
In the estimate of the lunar radial expansion, we consider the effect of melting as well as thermal expansion/contraction; the radius change $\Delta R$ is given by
\begin{equation}
 \label{dr_ch}
      \Delta R = \frac{1}{r_{\mathrm{p}}} \left[ \frac{\alpha_{\mathrm{c}} \Delta T_{\mathrm{c}} }{2} r_{\mathrm{c}}^2 + \int_{r_{\mathrm{c}}}^{r_{\mathrm{p}}}\int_{0}^{\pi}
     \left( \alpha_{\mathrm{m}} \Delta T +  \frac{\Delta v_{\mathrm{l}}}{v_{\mathrm{0}}} \Delta \phi \right) r dr d\theta\right] \,.
   \end{equation}
   
   In this equation, the first and second terms in the bracket of the right-hand side represent the volume changes of the core and the mantle, respectively.
   Here, $\alpha_\mathrm{c}$ is the thermal expansivity of the core, $\Delta T_\mathrm{c}\equiv{T_\mathrm{c}(t)-T_\mathrm{c}(t=0)}$ is the deviation of the temperature $T_\mathrm{c}$ in the core at the elapsed time $t$ from its initial value $T_\mathrm{c}(t=0)$, and $\Delta T_\mathrm{mid}\equiv{T(r,\theta,t)-T(r,\theta,t=0)}$  and $\Delta \phi\equiv{\phi(r,\theta,t)-\phi(r,\theta,t=0)}$ are the deviations in the temperature $T$ and melt-content $\phi$ in the mantle from their initial values, respectively.
   We assume that the initial condition in this model is the state of immediately after mantle overturn (see below), based on that the tectonic features of expansion/contraction in the crust are not recorded during the solidification of the magma ocean \cite{Elkins&Bercovici}.
  
 \subsection{The initial condition}
 \label{The initial condition}
\begin{figure}
\noindent\includegraphics[scale=0.6]{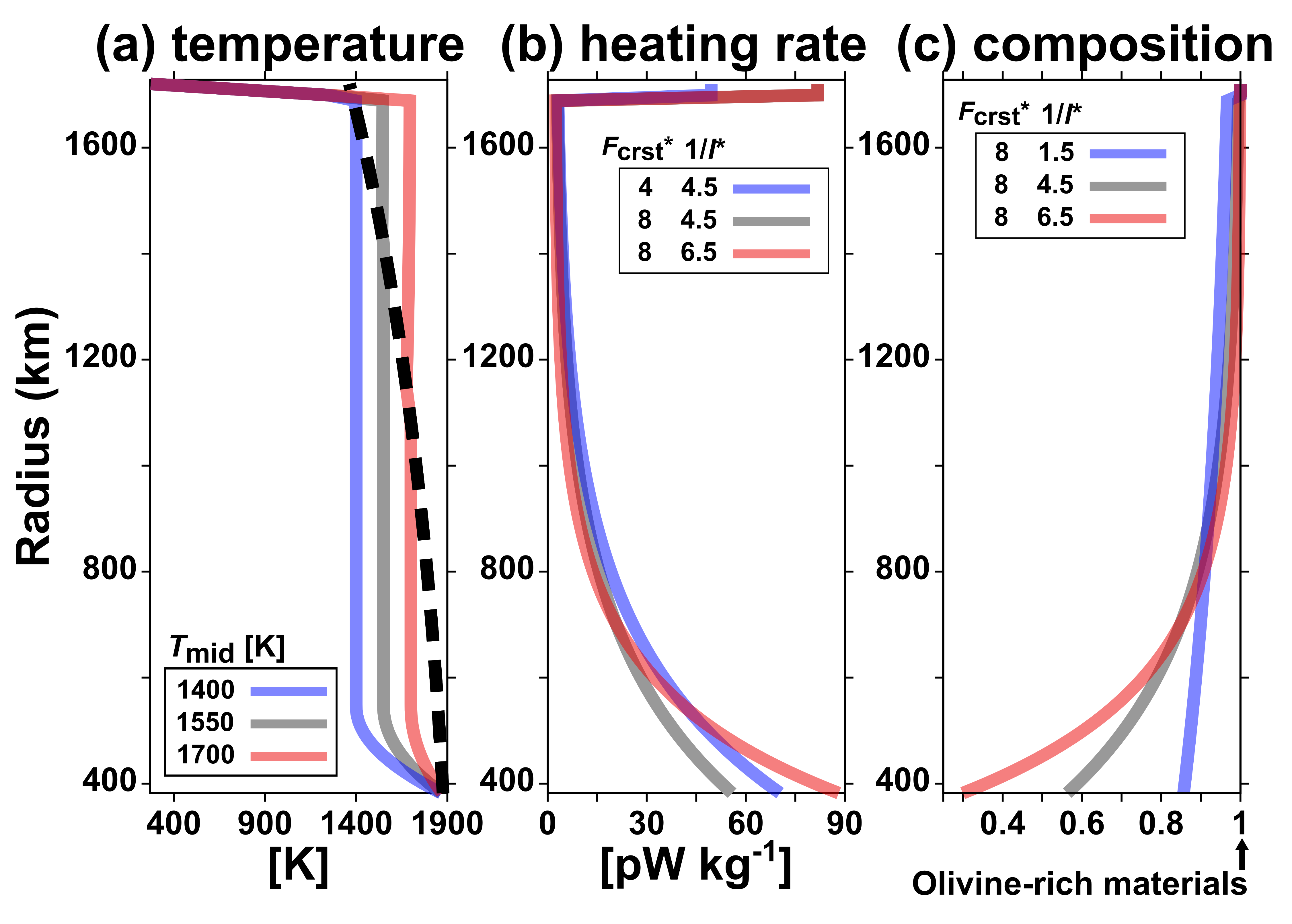}
\caption{An illustration of the initial distributions of (a) temperature, (b) internal heating rate, and (c) bulk composition. In (a), the black dashed line is the solidus which is calculated by Eq. \ref{mt_eq}. The meaning of variable parameters $T_{\mathrm{mid}}$, $l^*$, $F_{\mathrm{crst}}^*$ are described in Table \ref{free-p}.}
\label{initial}
\end{figure}

  \noindent The initial thermo-chemical state of the mantle is specified by the initial distributions of heating rate $q$, bulk composition $\xi_{\mathrm{b}}$, and temperature $T$. Here, the bulk composition is calculated form the composition of solid-phase $\xi_\mathrm{s}$ and that of $\xi_\mathrm{l}$
   \begin{equation}
   \xi_{\mathrm{b}} = \left( 1- \phi \right) \xi_{\mathrm{s}} + \phi \xi_{\mathrm{l}} \,.
   \end{equation}
   Fig. \ref{initial} shows an example of the initial distributions of internal heating rate, composition, and temperature. This initial condition is motivated by earlier models of the mantle overturn that is expected to have occurred after the solidification of the magma ocean \cite<e.g.,>{snyder,alley,bouk,rapp}. In the last phase of crystal fractionation of the magma ocean, there remains a dense layer of the IBC with the concentration of urKREEP (K, rare earth elements, and P-rich material) at the top of the mantle. This layer is expected to sink down due to the gravitational instability at the base of the mantle \cite<e.g.,>{R&K,hess&P,deVries}.

The initial distribution of the temperature is obtained from that of the `reduced’ enthalpy (see Eq. \ref{enthalpy}). The initial distribution of the `reduced’ enthalpy is 
       \begin{equation}
       h = \mathrm{min} \left( h_{\mathrm{sur}}, h_{\mathrm{mantle}} \right)  \,,
   \end{equation}
where
       \begin{equation}
       h_{\mathrm{sur}} = C{_{p}} \left[ T_{\mathrm{sur}} + \delta_\mathrm{crst} \left( 1 - \frac{r}{r_p}  \right) \right] \,,
   \end{equation}

\begin{eqnarray}
h_{\mathrm{mantle}} =
\left\{
\begin{array}{cl}
C{_{p}} T_{\mathrm{mid}} & \mathrm{if} \; r > r_{\mathrm{l}} \\
C{_{p}} \left[ T_{\mathrm{mid}} + \left( T_{\mathrm{c}} - T_{\mathrm{mid}} \right) \left(\displaystyle \frac{r_{\mathrm{l}} - r}{r_{\mathrm{l}} - r_{\mathrm{c}}}\right)^2 \right] & \mathrm{if} \; r < r_{\mathrm{l}}
\end{array}
\right. \,.
\end{eqnarray}
Here, $\delta_\mathrm{crst}$ and $r_{\mathrm{l}}$ are the constants arbitrarily chosen to be $\delta_\mathrm{crst} = 79.5 \times 10^3$ K to adopt the crust about 35 km thickness and
%930, 37 km
$r_{\mathrm{l}} = 550$ km which is determined by assuming the post-overturn stratification, respectively \cite<e.g.,>{bouk,Mitchell2021}. 
$T_{\mathrm{mid}}$ is the initial temperature in the mid-mantle;
$T_{\mathrm{c}} = 1875 \; \mathrm{K}$ the initial temperature of the core.
%1872.98 K
The value of $T_{\mathrm{c}}$ is based on the assumption that the temperature of the core is hotter than the mantle at 4.4 Gyr ago \cite{alley,bouk,Morbidelli}. We estimated the initial distribution of internal heat source $q =  q_{\mathrm{init}} $ in the mantle as  $\int_{r_{\mathrm{c}}}^{r_{\mathrm{p}}} q_\mathrm{init} r dr = \int_{r_{\mathrm{c}}}^{r_{\mathrm{p}}} q_0 r dr$, where $q_0 = 14.70 $ $\mathrm{pW \, kg^{-1}}$ is the average value of internal heating rate at the initial state of 4.4 Gyr ago estimated from Table 2 in \citeA{u2022}.
%This dimensionless value is $q_{\mathrm{init}}^* = q_{\mathrm{init}} / \left( \Delta h\frac{\kappa}{L^2} \right) $. 
We approximate the distribution of internal heating rate in the mantle discribed as
       \begin{equation}
       q_{\mathrm{init}} = q_{\mathrm{m}} + \Delta q \exp \left[ - \frac{r - r_\mathrm{c}}{l} \right]  \,,
   \end{equation}
  where $q_{\mathrm{m}} = 2.77 \; \mathrm{pW} \, \mathrm{kg}^{-1}$ is the internal heating rate of the olivine-rich materials in the mantle before mantle overturn \cite{yu}; $l$ the thickness of the basal layer which is enriched in the overturned materials; $\Delta q$ is calculated from the assumption that the excess internal heating rate in the urKREEP $Q_{\mathrm{ur}}$ defined by:
  \begin{equation}
      Q_{\mathrm{ur}} = \int_{r_{\mathrm{c}}}^{r_{\mathrm{crst}}}
      \Delta q \exp \left( \frac{r - r_\mathrm{c}}{l} \right)  r dr  \,,
   \end{equation}
  which value satisfies,
    \begin{equation}
      Q_{\mathrm{ur}} = \int_{r_{\mathrm{c}}}^{r_{\mathrm{p}}}
       q_0 r dr - \left( \int_{r_{\mathrm{crst}}}^{r_{\mathrm{p}}}
       q_\mathrm{crst}  r dr +\int_{r_{\mathrm{c}}}^{r_{\mathrm{crst}}}
      \Delta q_\mathrm{m}  r dr \right) \,.
   \end{equation}
   Here, $q_\mathrm{crst}$ is the average initial heating rate of the crust which holds on the non-overturned urKREEP \cite{yu,zhao2019}. We assume that the value of $q_\mathrm{crst}$ is obtained from the ratio of the concentration of the HPEs in the crust to that in the mantle written as
   \begin{equation}
   F_{\mathrm{crst}}^* =
   \frac{\int_{r_{\mathrm{crst}}}^{r_{\mathrm{p}}}
       q_\mathrm{crst}  r dr}{ \int_{r_{\mathrm{c}}}^{r_{\mathrm{p}}}
       q_0 r dr - \int_{r_{\mathrm{crst}}}^{r_{\mathrm{p}}}
       q_\mathrm{crst}  r dr}  \,.
     \end{equation}
       Furthermore, we calculated the initial distribution of the bulk composition $\xi_{\mathrm{b}}$ in the mantle using the assumption of that the overturned materials with the composition $\left( \xi_{\mathrm{e}}=0.1 \right)$ are 7.5 times more enriched in HPEs than the bulk Moon $q_0$ \cite{hess&P} as
       \begin{equation}
      \xi_{\mathrm{b}} = 1- \frac{\Delta q \exp \left[ - \left( r - r_{\mathrm{c}}  \right)/ l  \right]}{7.5 q_{0}}  \left( 1- \xi_{\mathrm{e}}\right)\,,
   \end{equation}
   where in the crust, we assumed $\xi_{\mathrm{b}}=1$.

 \subsection{The parameter values}
 \label{The parameter values}
 \noindent We carried out numerical experiments at various values of the non-dimensional reference permeability $M^* \equiv \frac{k_{\mathrm{\phi_{\mathrm{0}}}} \rho_0 {g_{\mathrm{sur}}} L}{\kappa \eta_{\mathrm{melt}}}$ (see Eq. \ref{M_num}) and the Rayleigh number $Ra \equiv \frac{\rho_0 \alpha_{\mathrm{m}} \left( \Delta h / C_p \right) {g_\mathrm{sur}} L^3}{\kappa \eta_{0}}$ (see Eq. \ref{Ra_num}), as well as the sensitivity of viscosity to temperature $E_T^* = E_T \Delta h / C_p $, the initial temperature in the mid-mantle $T_\mathrm{mid}$, the thickness of the overturned layer $l^* = l/L$, and the ratio of the concentration of HPEs in the crust to that in the mantle $F_{\mathrm{crst}}^*$, where $L = r_\mathrm{p} - r_\mathrm{c}$, is the depth of the mantle.  The non-dimensional reference permeability $M^*$  is proportional to $k_{\mathrm{\phi_{\mathrm{0}}}}/ \eta_{\mathrm{melt}}$ and expresses how easily the upward migration of magma takes place. We varied the values of $M^\ast$ in the range of $5\le{M^\ast}\le100$, which corresponds to the change in the viscosity $\eta_{\mathrm{melt}}$ of magma by a factor of 20 (1 to 20 Pa s). 
 On the other hand, the Rayleigh number $Ra$ is inversely proportional to the reference viscosity and expresses how readily mantle convection occurs.
 We assume that this value is in the range of $2.15 \; \times 10^4$ to $10^6$, which is estimated from a typical value of viscosity for mantle materials \cite{karato&wu}. 
 %This value range of $Ra$ is near the critical Rayleigh number for the thermal convection of a Newtonian fluid with temperature-dependent viscosity in a bottom-heated plane layer (see Figure 8 of \cite{yanagisawa}).

Other parameters listed in Table \ref{free-p} are also varied to see how the numerical results depend on the thermal and compositional structure of the mantle assumed in the initial conditions. 
%There are three free parameters that define the initial conditions: the initial temperature in the mid-mantle $T_\mathrm{mid}$, the normalized thickness of the overturned layer $l^*=l/L$, and the ratio of the concentration of the HPEs in the crust to that in the mantle ${F_{\mathrm{crst}}^*}$. These parameters in the initial conditions are determined by previous models.
In the range of $T_\mathrm{mid}$ is decided by that of initial temperature in \citeA{Laneuville2013} and that of post-overturn temperature in \citeA{bouk}. ${l^*}$ is estimated from the density of the basal layer which is enriched in the overturned materials obtained by the results of previous overturn models \cite<e.g.,>{li2019,zhao2019}. The thickness of the overturned layer is thinner and more concentrated in HPEs and the IBC component with a lower ${l^*}$. We also assume in the value of ${F_{\mathrm{crst}}^*}$ from that estimated in \citeA{k&s} and \citeA{Spohn}.
   
 \begin{table}
 \caption{The variable parameters of the numerical models.}
 \label{free-p}
 \centering
 \begin{tabular}{l l l}
 \hline
  Variable parameter  & Meaning  & Range of value \\
   \hline
     $M^*$  & Dimensionless reference permeability  & $5 \;\mathrm{to} \; 100$   \\
    $Ra$  & Rayleigh number  & $2.15 \times 10^{4}$ to $10^{6}$\\
  $E_T^*$  & Dimensionless sensitivity of viscosity to temperature  & $3 \;\mathrm{to} \; 9$   \\
   $T_{\mathrm{mid}} $  & Initial temperature in the mid-mantle  & $1400$ to $1700$ K   \\
      $l^*$  & The thickness of the overturned layer after mantle overturn  & $1/6.5$ to $1/1.5$  \\
   $F_{\mathrm{crst}}^*$  & Concentration ratio of HPEs in the crust to the mantle  & $4$ to $32$  \\
 \hline
% \multicolumn{2}{l}{$^{a}$Footnote text here.}
 \end{tabular}
 \end{table}
 
  %%%%%%
  \section{Results} 
  \subsection{The reference case} 
 \noindent Figures \ref{for_journal}-\ref{distribution} show the reference case of Case Ref calculated at the reference permeability $M^* = 100$; the Rayleigh number $Ra= 2.15 \times 10^{6}$, corresponding to $\eta_0 = 10^{20}$ Pa s; the sensitivity of viscosity on temperature $E_T^* = 6$; the initial temperature at depth $T_{\mathrm{mid}}= 1550$ K; the initial thickness of the overturned layer $l^*=4.5$; the initial crustal fraction of the HPEs $F_{\mathrm{crst}}^*=8$ (Table \ref{free-p}).

\begin{sidewaystable}
 \caption{The values of the parameters listed in Table\ref{free-p} and results.}
 \label{case}
 \centering
 \begin{tabular}{l l l l l l l l l l l}
 \hline
  Case $\#$& $M^*$ & $Ra$ & $E_T^*$ & $T_{\mathrm{mid}} $ [K] & $1/l^*$ & $F_{\mathrm{crst}}^*$ & $\Delta R_{\mathrm{peak}}$ [km]& $t_{\mathrm{\Delta R}}$ [Gyr] & $\chi_{Gyr}$ [km $\mathrm{Gyr^{-1}}$] & $L_{\phi}$ [km]\\
 \hline
   Ref & 100 & $2.15 \times 10^{6}$ & 6 & 1550 & 4.5 & 8 & 3.00 & 0.66 & -1.01 & 25\\
   
   M50 & 50 & $2.15 \times 10^{6}$ & 6 & 1550 & 4.5 & 8 & 4.44 & 0.70 & -1.33 & 185\\
    M20 & 20 & $2.15 \times 10^{6}$ & 6 & 1550 & 4.5 & 8 & 6.62 & 0.89 & -1.47 & 225\\
    M5 & 5 & $2.15 \times 10^{6}$ & 6 & 1550 & 4.5 & 8 & 8.95 & 1.73 & -1.78 & 200\\

   Ra2.15e4 & 100 & $2.15 \times 10^{4}$ & 6 & 1550 & 4.5 & 8 & 3.84 & 1.04 & -1.16 & 75\\
  Ra2.15e5 & 100 & $2.15 \times 10^{5}$ & 6 & 1550 & 4.5 & 8 & 3.80 & 1.05 & -1.17 & 60\\
 Ra7.15e5 & 100 & $7.15 \times 10^{5}$ & 6 & 1550 & 4.5 & 8 & 3.53 & 0.83 & -0.96  & 40\\

 ET3 & 100 & $2.15 \times 10^{6}$ & 3 & 1550 & 4.5 & 8 & 3.06 & 0.78 & -0.76 & 15\\
  ET9 & 100 & $2.15 \times 10^{6}$ & 9 & 1550 & 4.5 & 8 & 3.30 & 0.66 & -1.28 & 165\\

   Tm1400 & 100 & $2.15 \times 10^{6}$ & 6 & 1400 & 4.5 & 8 & 3.62 & 1.36 & -0.57 & 30\\  
   Tm1475 & 100 & $2.15 \times 10^{6}$ & 6 & 1475 & 4.5 & 8 & 3.34 & 1.13 & -0.67 & 25\\  
   Tm1625 & 100 & $2.15 \times 10^{6}$ & 6 & 1625 & 4.5 & 8 & 3.80 & 0.50 & -1.00 & 30\\  
   Tm1700 & 100 & $2.15 \times 10^{6}$ & 6 & 1700 & 4.5 & 8 & 4.23 & 0.42 & -1.41 & 115\\   

   l1.5 & 100 & $2.15 \times 10^{6}$ & 6 & 1550 & 1.5 & 8 & 3.01 & 1.04 & -1.17 & 190\\
   l2.5 & 100 & $2.15 \times 10^{6}$ & 6 & 1550 & 2.5 & 8 & 3.66 & 0.86 & -1.43 & 80\\
   l3.5 & 100 & $2.15 \times 10^{6}$ & 6 & 1550 & 3.5 & 8 & 3.51 & 0.69 & -0.97 & 170\\
   l5.5 & 100 & $2.15 \times 10^{6}$ & 6 & 1550 & 5.5 & 8 & 2.92 & 0.85 & -1.20 & 20\\
   l6.5 & 100 & $2.15 \times 10^{6}$ & 6 & 1550 & 6.5 & 8 & 2.50 & 0.62 & -0.93 & 15\\
   l0 & 100 & $2.15 \times 10^{6}$ & 6 & 1550 & 0 & 8 & 0.22 & 1.45 & -0.62 & 215\\

   Fcrst4 & 100 & $2.15 \times 10^{6}$ & 6 & 1550 & 4.5 & 4 & 3.85 & 0.69 & -1.23 & 25\\
   Fcrst16 & 100 & $2.15 \times 10^{6}$ & 6 & 1550 & 4.5 & 16 & 1.69 & 1.07 & -1.03 & 190\\
   Fcrst32 & 100 & $2.15 \times 10^{6}$ & 6 & 1550 & 4.5 & 32 & -0.50 & 1.46 & -0.95 & 395\\

   M50-Ra2.15e4 & 50 & $2.15 \times 10^{4}$ & 6 & 1550 & 4.5 & 8 & 5.25 & 0.79 & -0.83 & 205\\
   M20-Ra2.15e4 & 20 & $2.15 \times 10^{4}$ & 6 & 1550 & 4.5 & 8 & 8.91 & 1.35 & -0.82 & 305\\
   M5-Ra2.15e4 & 5 & $2.15 \times 10^{4}$ & 6 & 1550 & 4.5 & 8 & 12.21 & 3.36 & -0.37 & 285\\
   M0-Ra2.15e4 & 0 & $2.15 \times 10^{4}$ & 6 & 1550 & 4.5 & 8 & 15.32 & 4.40 & 1.21 & 375\\

   M50-l5.5 & 50 & $2.15 \times 10^{6}$ & 6 & 1550 & 5.5 & 8 & 4.00 & 0.73 & -1.22 & 160\\
   M20-l5.5 & 20 & $2.15 \times 10^{6}$ & 6 & 1550 & 5.5 & 8 & 7.09 & 0.83 & -1.12 & 210\\
   M5-l5.5 & 5 & $2.15 \times 10^{6}$ & 6 & 1550 & 5.5 & 8 & 9.00 & 1.80 & -1.25 & 195\\

    Tm1400-l5.5 & 100 & $2.15 \times 10^{6}$ & 6 & 1400 & 5.5 & 8 & 3.86 & 1.12 & -0.10 & 155\\
    Tm1700-l5.5 & 100 & $2.15 \times 10^{6}$ & 6 & 1700 & 5.5 & 8 & 4.09 & 0.35 & -1.16 & 40\\

    Tm1400-l1.5 & 100 & $2.15 \times 10^{6}$ & 6 & 1400 & 1.5 & 8 & 4.91 & 1.68 & -0.84 & 210\\
    Tm1700-l1.5 & 100 & $2.15 \times 10^{6}$ & 6 & 1700 & 1.5 & 8 & -0.21 & 0.53 & -1.08 & 105\\
    Tm1400-l5.5 & 5 & $2.15 \times 10^{6}$ & 6 & 1400 & 5.5 & 8 & 9.83 & 3.40 & -0.13 & 270\\

    Ra2.15e4-l5.5 & 100 & $2.15 \times 10^{4}$ & 6 & 1550 & 5.5 & 8 & 3.44 & 1.09 & -1.17 & 25\\
    Ra2.15e5-l5.5 & 100 & $2.15 \times 10^{5}$ & 6 & 1550 & 5.5 & 8 & 3.15 & 1.10 & -1.17 & 25\\
    M50-Ra4-l5.5 & 50 & $2.15 \times 10^{4}$ & 6 & 1550 & 5.5 & 8 & 4.98 & 1.15 & -1.17 & 175\\
    M20-Ra4-l5.5 & 20 & $2.15 \times 10^{4}$ & 6 & 1550 & 5.5 & 8 & 9.43 & 1.62 & -0.91 & 310\\
    M5-Ra4-l5.5 & 5 & $2.15 \times 10^{4}$ & 6 & 1550 & 5.5 & 8 & 12.57 & 3.19 & -0.37 & 290\\
    M5-Ra5-l5.5 & 5 & $2.15 \times 10^{5}$ & 6 & 1550 & 5.5 & 8 & 12.46 & 3.02 & -0.19 & 285\\
 \hline
% \multicolumn{2}{l}{$^{a}$Footnote text here.}
 \end{tabular}
 \end{sidewaystable}
%% \begin{abstract} starts the second page
%
\setcounter{table}{2}
\begin{sidewaystable}
 \caption{Continued.}
 \centering
 \begin{tabular}{l l l l l l l l l l l}
 \hline
  Case $\#$& $M^*$ & $Ra$ & $E_T^*$ & $T_{\mathrm{mid}} $ [K] & $1/l^*$ & $F_{\mathrm{crst}}^*$ & $\Delta R_{\mathrm{peak}}$ [km]& $t_{\mathrm{\Delta R}}$ [Gyr] & $\chi_{Gyr}$ [km $\mathrm{Gyr^{-1}}$] & $L_{\phi}$ [km]\\
 \hline
   ET9-l5.5 & 100 & $2.15 \times 10^{6}$ & 9 & 1550 & 5.5 & 8 & 3.28 & 0.60 & -0.87 & 85\\
    l5.5-Fcrst16 & 100 & $2.15 \times 10^{6}$ & 6 & 1550 & 5.5 & 16 & 1.74 & 0.890 & -0.85 & 185\\
    l5.5-Fcrst32 & 100 & $2.15 \times 10^{6}$ & 6 & 1550 & 5.5 & 32 & -0.21 & 1.30 & -0.91 & 315\\
    
   ET3-l5.5 & 100 & $2.15 \times 10^{6}$ & 3 & 1550 & 5.5 & 8 & 2.99 & 0.80 & -1.03 & 20\\
   ET3-l1.5 & 100 & $2.15 \times 10^{6}$ & 3 & 1550 & 1.5 & 8 & 4.00 & 1.11 & -0.82 & 160\\ 
   M5-ET3-l5.5 & 5 & $2.15 \times 10^{6}$ & 3 & 1550 & 5.5 & 8 & 10.40 & 2.34 & -2.00 & 180\\ 
   Ra4-ET3-l5.5 & 100 & $2.15 \times 10^{4}$ & 3 & 1550 & 5.5 & 8 & 3.88 & 1.12 & -1.67 & 25\\ 
   Ra4-ET9-l5.5 & 100 & $2.15 \times 10^{4}$ & 9 & 1550 & 5.5 & 8 & 3.00 & 0.66 & -1.17 & 150\\ 
   ET9-l1.5 & 100 & $2.15 \times 10^{6}$ & 9 & 1550 & 1.5 & 8 & 2.19 & 0.85 & -1.39 & 160\\ 
   M5-ET9-l5.5 & 5 & $2.15 \times 10^{6}$ & 9 & 1550 & 5.5 & 8 & 8.75 & 1.62 & -1.91 & 185\\ 

   Tm14-l5.5-Fcrst16 & 100 & $2.15 \times 10^{6}$ & 6 & 1400 & 5.5 & 16 & 2.46 & 1.06 & 0.07 & 225\\ 
   Tm17-l5.5-Fcrst16 & 100 & $2.15 \times 10^{6}$ & 6 & 1700 & 5.5 & 16 & 2.93 & 0.45 & -1.10 & 50\\
   ET3-l5.5-Fcrst16 & 100 & $2.15 \times 10^{6}$ & 3 & 1550 & 5.5 & 16 & 1.96 & 0.83 & -0.94 & 50\\ 
   ET9-l5.5-Fcrst16 & 100 & $2.15 \times 10^{6}$ & 9 & 1550 & 5.5 & 16 & 1.80 & 0.80 & -1.27 & 195\\
   Ra4-l5.5-Fcrst16 & 100 & $2.15 \times 10^{4}$ & 6 & 1550 & 5.5 & 16 & 2.14 & 1.46 & -1.10 & 100\\ 
   Ra5-l5.5-Fcrst16 & 100 & $2.15 \times 10^{5}$ & 6 & 1550 & 5.5 & 16 & 2.06 & 1.58 & -0.86 & 115\\ 
   l1.5-Fcrst16 & 100 & $2.15 \times 10^{6}$ & 6 & 1550 & 1.5 & 16 & 0.83 & 1.45 & -0.97 & 90\\ 
   l3.5-Fcrst16 & 100 & $2.15 \times 10^{6}$ & 6 & 1550 & 3.5 & 16 & 1.88 & 0.98 & -0.69 & 185\\ 
   M5-l5.5-Fcrst16 & 5 & $2.15 \times 10^{6}$ & 6 & 1550 & 5.5 & 16 & 4.37 & 1.79 & -0.71 & 240\\ 
   M20-l5.5-Fcrst16 & 20 & $2.15 \times 10^{6}$ & 6 & 1550 & 5.5 & 16 & 3.63 & 1.07 & -0.60 & 260\\ 
   M50-l5.5-Fcrst16 & 50 & $2.15 \times 10^{6}$ & 6 & 1550 & 5.5 & 16 & 2.65 & 0.80 & -1.05 & 205\\ 

   No-conv-HPEs-Mass-tr & 100 & $2.15 \times 10^{2}$ & 6 & 1550 & 4.5 & 8 & -0.65 & 1.11 & -0.71 & 195\\
   No-HPEs-Mass-tr & 100 & $2.15 \times 10^{6}$ & 6 & 1550 & 4.5 & 8 & -4.12 & 0.49 & -0.80 & 20\\
   No-Mass-tr & 100 & $2.15 \times 10^{6}$ & 6 & 1550 & 4.5 & 8 & -5.01 & 0.55 & -1.30 & 15\\
   beta67-Ref ($\beta = 0.067$) & 100 & $2.15 \times 10^{6}$ & 6 & 1550 & 4.5 & 8 & -0.60 & 2.38 & -0.85 & 20\\
 \hline
\multicolumn{11}{l}{%
  \begin{minipage}{18.5cm}%
    $\Delta R_{\mathrm{peak}}$ and $t_{\mathrm{\Delta R}}$ stands for the magnitude and timing of the peak of radial expansion, respectively; $\chi_{Gyr}$ the contraction rate for the past 1 Gyr obtained by least-squares method; $L_{\phi}$ the shallowest depth levels of partially molten regions other than that in the initial state.  %
  \end{minipage}%
}\\
 \end{tabular}
 \end{sidewaystable}

\subsubsection{Thermal and structural evolution of the mantle}
\label{Thermal and structural evolution of the mantle}
  \begin{figure}
  \noindent\includegraphics[scale=0.8]{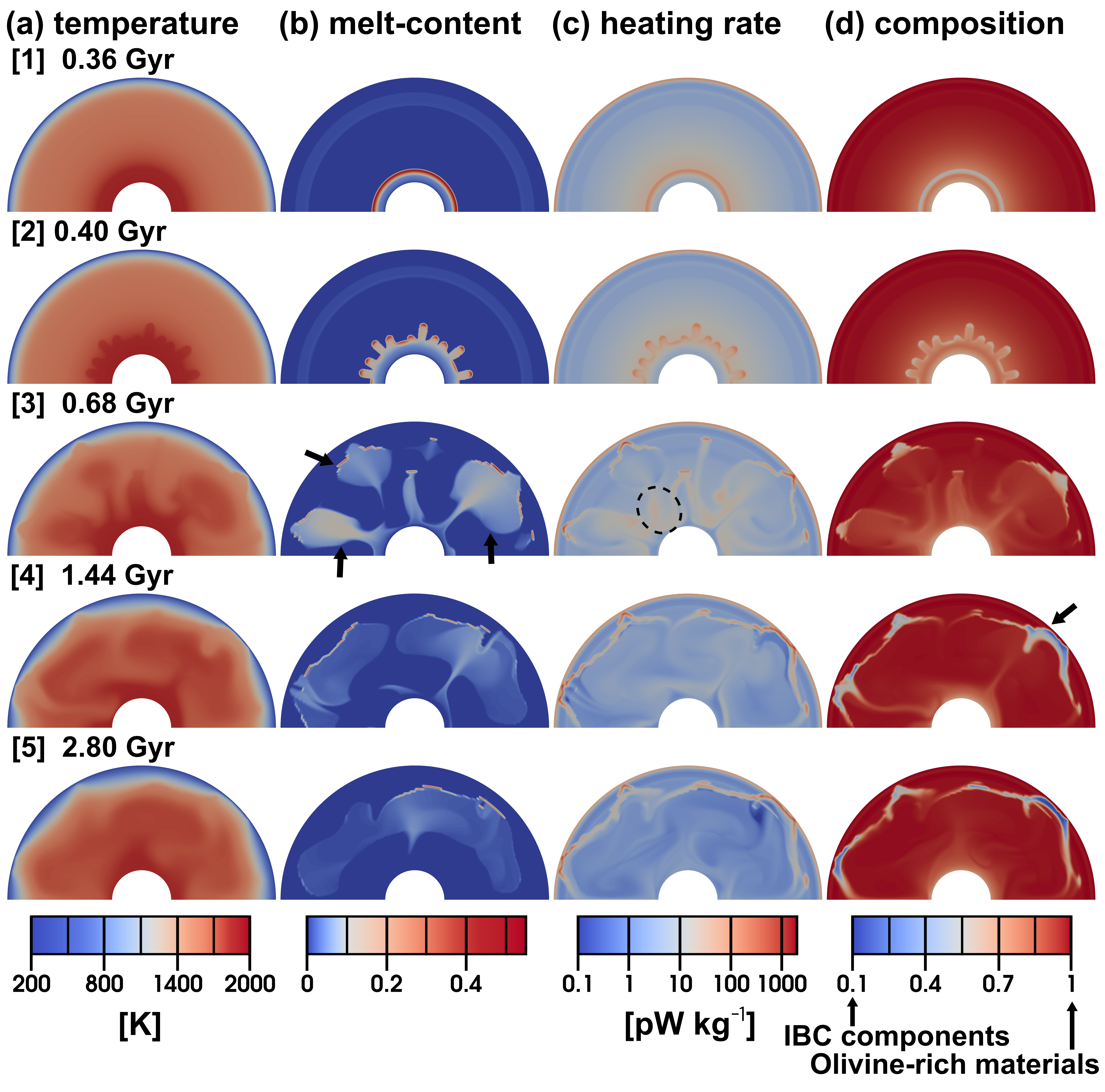}
 \caption{Snapshots of the distributions of (a) temperature $T$, (b) melt-content $\phi$, (c) internal heating rate $q$, and (d) composition $\xi_{\mathrm{b}}$ calculated for Case Ref where $M^* = 100$; $Ra= 2.15 \times 10^{6}$, corresponding to $\eta_0 = 10^{20}$ Pa s; $E_T^* = 3$; $T_{\mathrm{mid}}= 1550$ K; $l^*=4.5$; $F_{\mathrm{crst}}^*=8$. The elapsed times are indicated in the figure. In (d), the blue color stands for the IBC component, while the red color for the olivine-rich end-member. The numbers [1] to [5] correspond with those of Figure \ref{radius}.}
  \label{for_journal}
 \end{figure}
 
\begin{figure}
\noindent\includegraphics[scale=0.5]{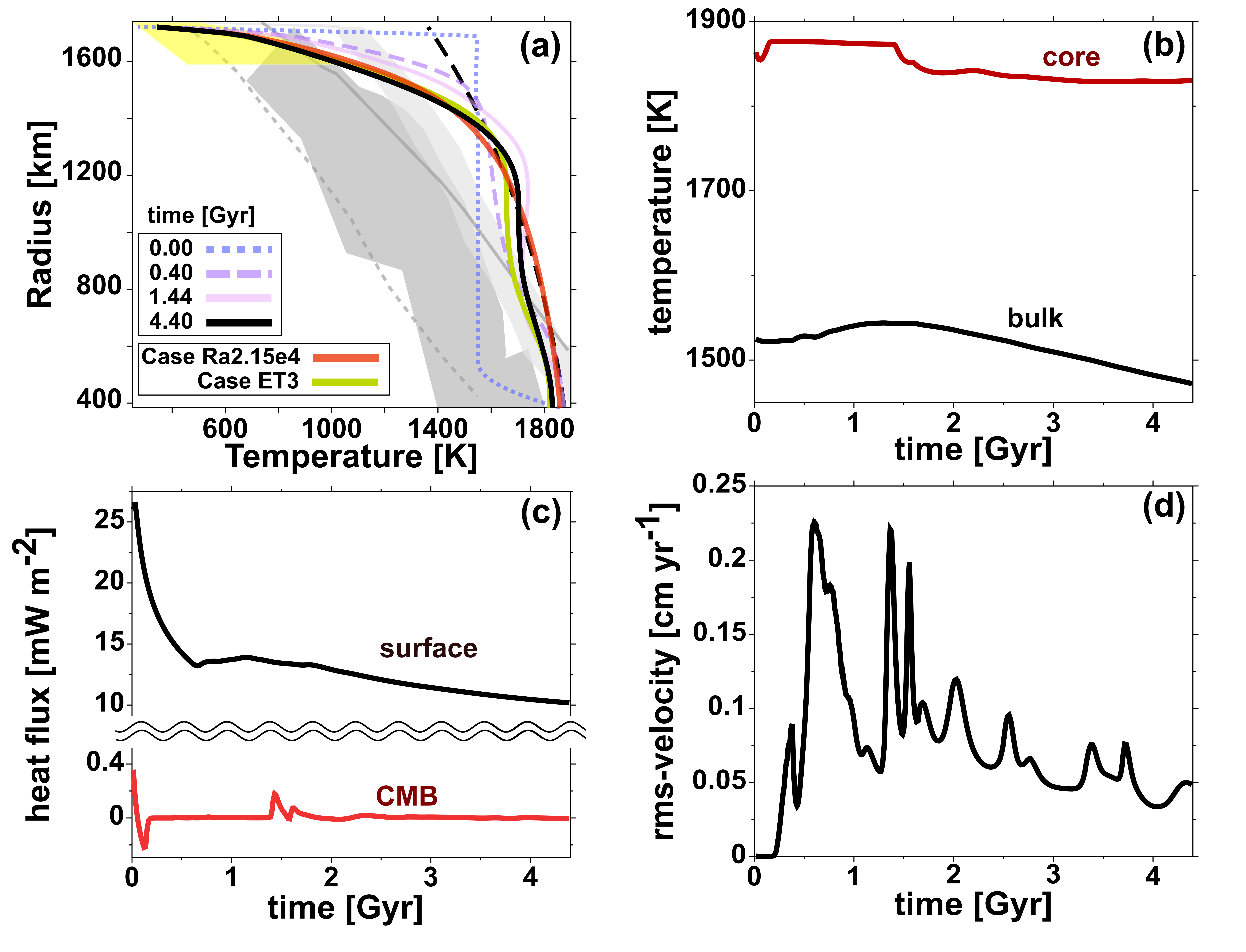}
\caption{(a) The horizontal averages of the temperature-distributions at various elapsed times for Case Ref. Also shown are the horizontally averaged temperature at 4.4 Gyr for Case Ra2.15e4 ($Ra$ = 2.15 $\times 10^4$) and ET3 ($E_T^*$ = 3). In (a), the gray and light-gray areas are the temperature distributions in today’s lunar mantle inferred by \citeA{khan2006} and \citeA{khan2014}, respectively; the yellow areas are those estimated from the heat flux at the surface \cite{Siegler2014,Siegler2022}; the gray lines are those estimated under the assumption that the mantle consists of dry olivine (the solid line) and wet olivine with 0.01 wt $\%$ $\mathrm{H_2O}$ (the dashed line) by \citeA{karato2013}. Also shown are (b) the average temperature in the mantle (bulk) and the temperature of the core; (c) the horizontal average of heat flux on the surface and the core-mantle boundary (CMB); (d) the root-mean-square average of matrix-velocity in the mantle all plotted against time.}
\label{distribution}
\end{figure}

  \noindent Figures \ref{for_journal} and \ref{distribution} as well as the animation in supporting materials (Movie S1-4) illustrate how the mantle evolves dynamically by magmatism and mantle convection. The partially molten region in the uppermost mantle assumed in the initial condition shrinks with time owing to conductive cooling from the surface boundary (Figure \ref{for_journal}b for 0.36 Gyr to 0.68 Gyr). In contrast, the temperature rises in the deep mantle that is enriched in HPEs in the initial condition, and magma is generated there within the first 150 Myr (Figures \ref{for_journal}a and \ref{distribution}a; see also Movie S1-3). 
  %The partially molten region in the deep mantle conducts heat to the core; thereby, the core temperature increases (Figure \ref{distribution}b, c). 
  The distributions of melt-content, HPEs, and bulk composition in the deep mantle are laterally uniform at the beginning of the calculated evolutionally history of the mantle (Figure \ref{for_journal}b-d for 0.36 Gyr). However, finger-like structures, or ``melt-fingers’’, eventually develop along the top of the partially molten region at a depth after around 0.4 Gyr. The fingertips are enriched in HPEs and the IBC component because of their transport by upward migrating magma (Figure \ref{for_journal} for 0.40 Gyr). Following the development of melt-fingers, partially molten plumes develop and ascend often along the fingers (see the arrows in Figure \ref{for_journal}b) to induces peaks in the plot of rms-velocity in Figure \ref{distribution}d. 
     Most of the partially molten plumes ascend to the uppermost mantle and reach the depth levels as shallow as around 25 km by 0.7 Gyr (Figure \ref{for_journal}b and Table \ref{case}). These plumes in the uppermost mantle gradually solidify as they are cooled from the surface boundary. Some partially molten plumes, however, solidify in the mid and deep mantle (see the dashed circle in Figure \ref{for_journal}c for 0.68 Gyr). The solidified materials enriched in HPEs and the IBC components then sink to the deep mantle because of their compositionally induced negative buoyancy (see Figure \ref{for_journal}d; see also Movie S4 for around 0.60 Gyr and 0.84 Gyr). Internal heating by HPEs in these materials induces further partial melting in the deep mantle. As a consequence, 
    partially molten plumes develop until around 4 Gyr, although they become fainter with time as the HPEs decline after around 2 Gyr (Figures \ref{for_journal}b and \ref{distribution}b-d).
  %I dont come up with what to write the rms
   Some materials in the uppermost mantle are enriched in HPEs and the IBC component by partially molten plumes (Figure \ref{for_journal}c, d). 
   These materials founder into the deep mantle owing to their compositionally induced negative buoyancy (see the arrow in Figure \ref{for_journal}d). The return flow of this foundering generates magma that is not enriched in HPEs and by decompression melting (Figure \ref{for_journal}b-d for 1.44 Gyr; see also Movie S2-4 for around 1.28 Gyr and 1.44 Gyr).
  % (The melt-fingers occur in the model where the numerical mesh turns 128 into 64 and 256 mesh points in the radial direction; therefore, it is not a product of computational instability.
   We will discuss more about melt-fingers and partially molten plumes in Section \ref{Melt-fingers and partially molten plumes}, below.
   
  \subsubsection{radius change}
  
\begin{figure}
\noindent\includegraphics[scale=0.6]{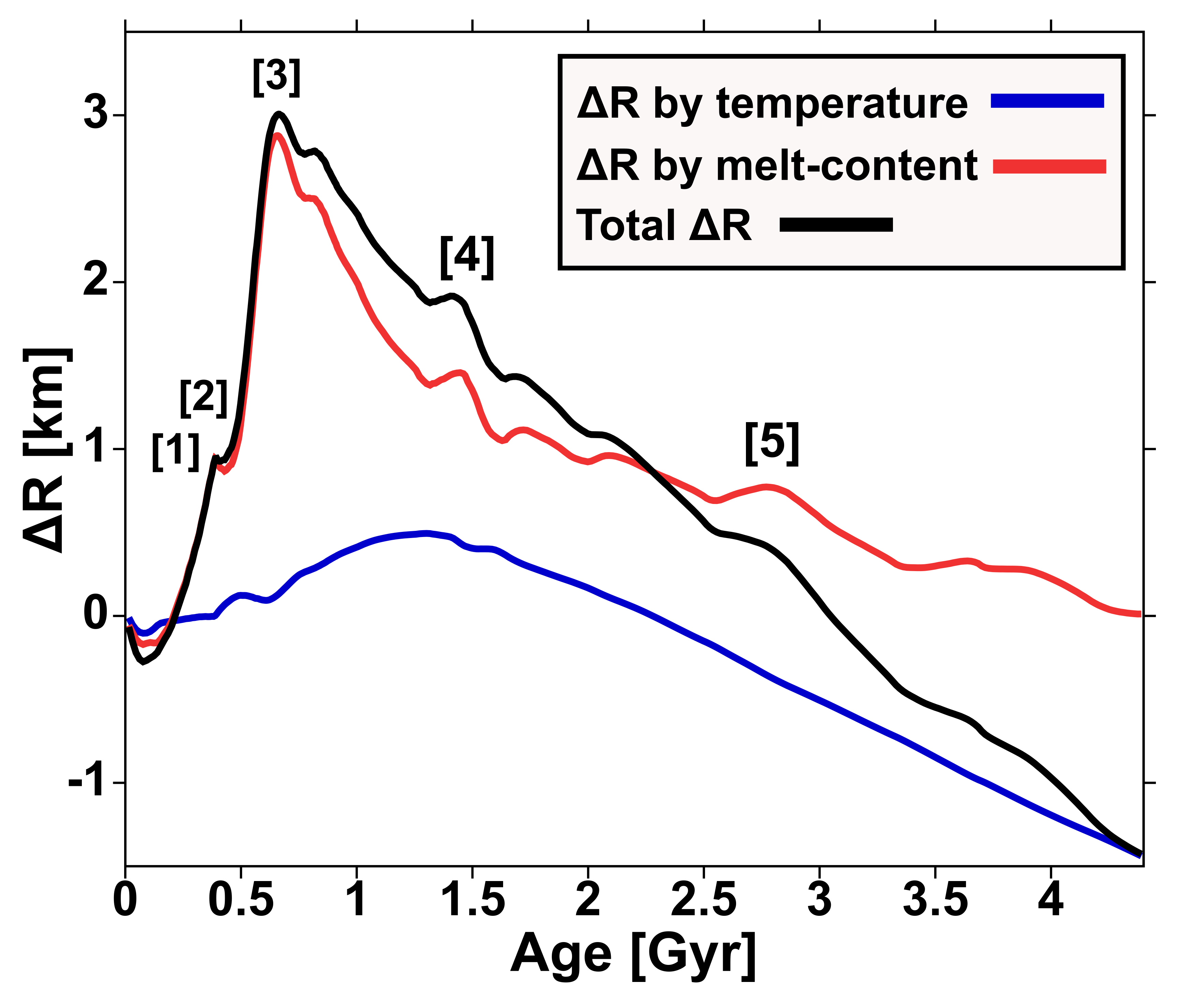}
\caption{The history of radius change of planet calculated in the reference case (Case Ref). The radius change $\Delta R$ is defined by Eq. \ref{dr_ch}; the blue and red lines indicate the contribution of thermal expansion/contraction and melting, respectively, to the total radius change (the black line). The numbers [1] to [5] correspond with those of Figure \ref{for_journal}.}
\label{radius}
\end{figure}
  
  \noindent Figure \ref{radius} shows how the planetary radius changes with time owing to the mantle evolution shown in Figures \ref{for_journal} and \ref{distribution}. The radius of a planet changes with time by two reasons, thermal expansion/contraction, and melting of the mantle (see Eq. \ref{dr_ch}), as indicated by the blue and red lines in the figure, respectively. 
  The blue line in Figure \ref{radius} shows that the planet thermally expands by 0.4 km in the early history as the mantle is heated up and then contracts until the end of the calculation as the mantle is cooled (Figure \ref{distribution}b). 
  On the other hand, the red line indicates the planet expands by 3 km for the first 0.7 Gyr owing to widespread partial melting of the mantle caused by melt-fingers and plumes (the peak [3] in Figure \ref{radius}). After that, the planet gradually contracts with time as the mantle solidifies. Note that the contraction is not monotonous: slight expansion occurs several times when partially molten plumes develop (see [4] and [5] in Figures \ref{for_journal} and \ref{radius}). 
  As a whole, the black line in Figure \ref{radius} indicates that the planet radially expands by about 3 km during the first 0.7 Gyr and then contracts at the rate of around $-1.0 \; \mathrm{km}\;\mathrm{Gyr^{-1}}$ until the end of the calculated history (Table \ref{case}).

 \subsubsection{Melt-fingers and partially molten plumes}
\label{Melt-fingers and partially molten plumes}

\begin{figure}
  \noindent\includegraphics[scale=0.7]{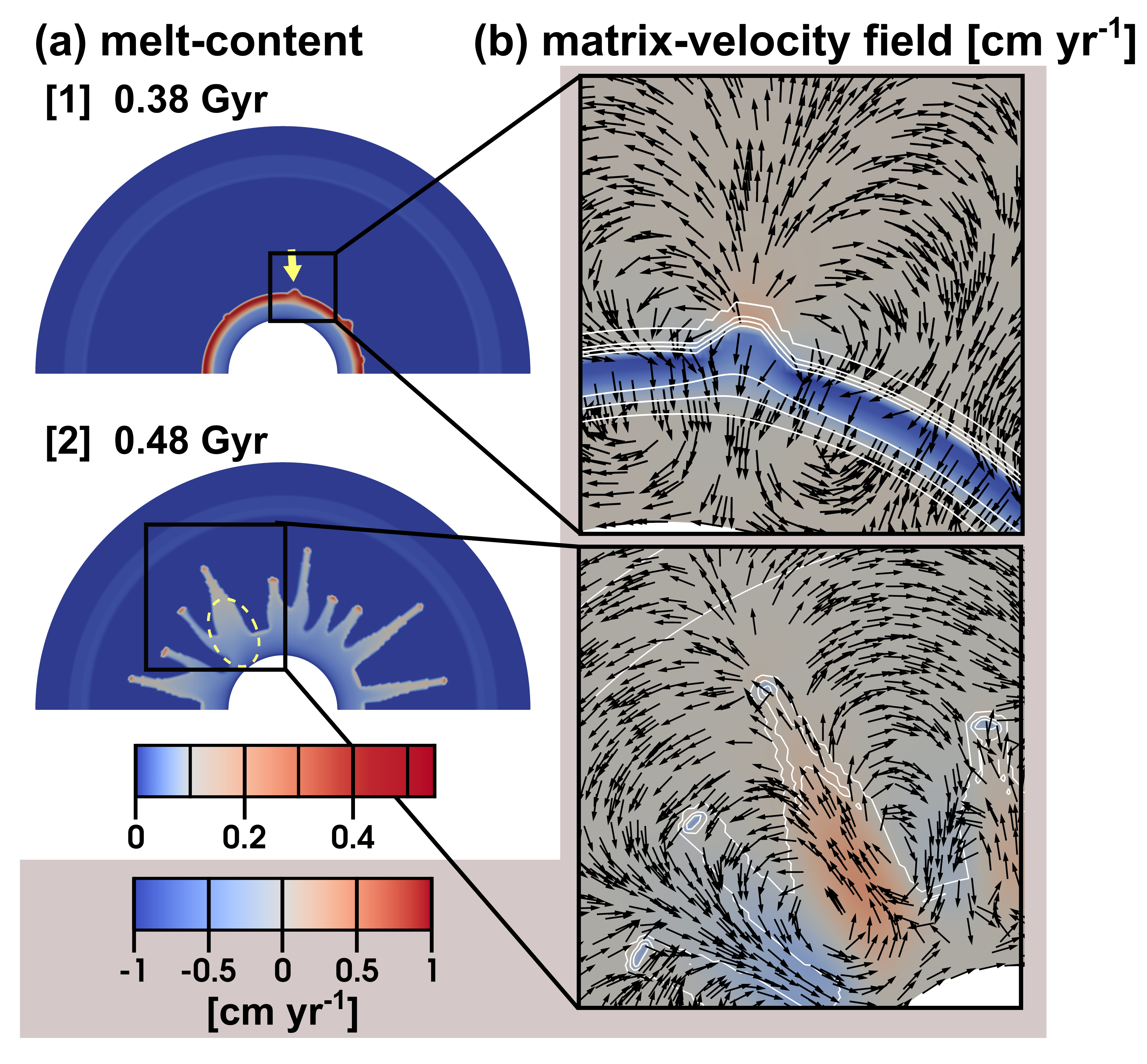}
  \caption{Snapshots of the (a) melt-content $\phi$ and (b) matrix-velocity field in the reference case (Case Ref). In (b), the red and blue colors show the regions where the convective flow points upward and downward, respectively; the arrows express the direction of convective flow but not its magnitude. The contour lines show the $\phi$-distribution with the contour interval of 0.1 starting from 0.}
\label{finger_and_plume}
  \end{figure}
   
   \noindent To understand why melt-fingers grow from the partially molten region at depths at around 0.4 Gyr in Figure \ref{for_journal}a, b, we delineate the velocity field of matrix around a nascent finger in Figure \ref{finger_and_plume}a (see the bump indicated by the yellow arrow). Figure \ref{finger_and_plume} for 0.38 Gyr shows that melt-fingers develop and grow upward owing to matrix-expansion around the fingertip caused by injection of upward migrating magma. Indeed, the matrix velocity around the fingertips diverges as indicated by the arrows in Figure \ref{finger_and_plume}b. (We recalculated melt-fingers on a mesh with twice higher resolution and confirmed that fingers still grow; melt-fingers are not an artifact of numerical instability.)
 
   After 100 Myr from the development of melt-fingers, partially molten plumes grow owing to the melt-buoyancy (see in Figure \ref{finger_and_plume} for 0.48 Gyr). The stem of melt-finger indicated by the yellow dashed circle in Figure \ref{finger_and_plume}a becomes thicker with time as the melt that migrates upward from the partially molten layer at depth accumulates. Because of the buoyancy of accumulated melt, both the matrix and melt migrate upward (Figure \ref{finger_and_plume}): the thickened area ascends as a plume at 0.48 Gyr. 

  \subsection{The occurrence of melt-fingers and partially molten plumes} 
\begin{figure}
\noindent\includegraphics[width=\textwidth]{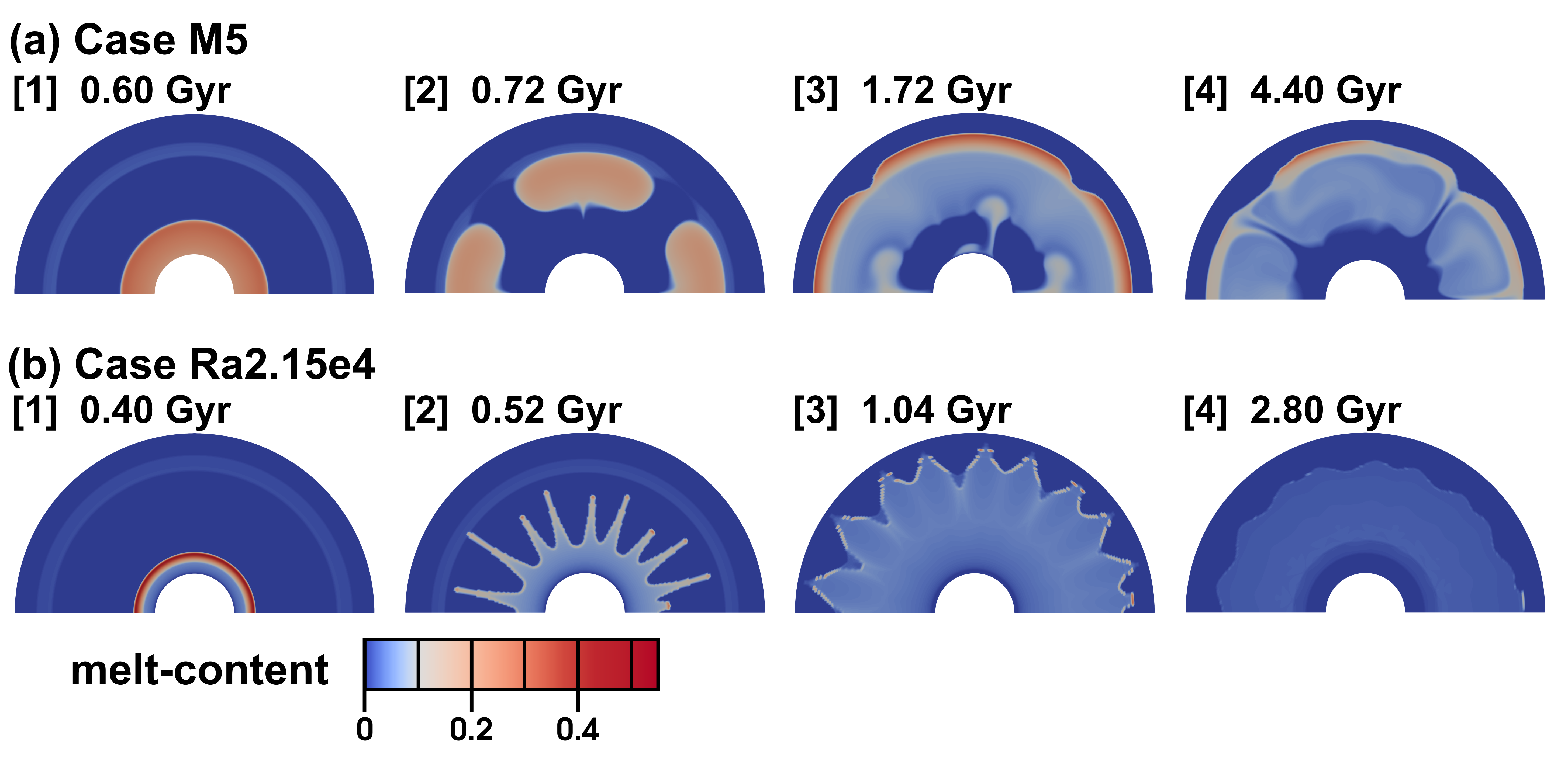}
\caption{Snapshots of the distributions of melt-content $\phi$ calculated in cases with (a) a reduced reference permeability (Case M5) and (b) a reduced Rayleigh number (Case Ra2.15e4). The numbers [1] to [5] correspond with those of Figure \ref{all_radius}a, e.}
\label{per5_e4}
\end{figure}
  
    \noindent  To understand under what condition melt-fingers and partially molten plumes observed in the reference case grow, we calculated the model at various values of the reference permeability $M^*$, and the Rayleigh number $Ra$.
   
   The reference permeability $M^*$ influences the growth of melt-fingers; the snapshots of Case M5 ($M^* = 5$ and other parameters fixed at their default values) illustrate that the melt-fingers do not occur at the lower reference permeability of this case (Figure \ref{per5_e4}a). Instead, partially molten plumes that are broader than those observed in the reference case rise to the depth levels of 200 km by around 1.7 Gyr (Figure \ref{per5_e4}a for 1.72 Gyr and Table \ref{case}). 
    The plumes laterally extend and form a continuous layer of partially molten materials (Figure \ref{per5_e4}a). The partially molten layer then solidifies upon cooling from the surface boundary; the layer, however, still remains in the mid-mantle at 4.4 Gyr (Figure \ref{per5_e4}a). We observed that melt-fingers grow only in the cases calculated at $M^* \geq 50$ (see Figure \ref{pngfiguresample1} in supporting materials).
    
     At $M^* = 100$, we also calculated several models at various values of $Ra$ in the range from $2.15 \times 10^2$ to $2.15 \times 10^6$ (Table \ref{case}); we found that melt-fingers develop regardless of the values of $Ra$, implying that buoyancy does not play any role in the growth of melt-fingers (Figures \ref{per5_e4}b and \ref{pngfiguresample2} in supporting materials).
     %(Note that mantle convection does not occur at $Ra = 2.15 \times 10^2$). 
     This result reinforces the above conclusion that melt-fingers develop because of matrix-expansion around the fingertip (Section \ref{Melt-fingers and partially molten plumes}). We further calculated a case (Case No-conv-HPEs-Mass-tr) with the partition coefficient of HPEs $D = 1$ in Eq. \ref{coeffiecient} (see Appendix) and a model that starts from a compositionally uniform mantle with the bulk composition $\xi_{\mathrm{b}}=\xi_{\mathrm{e}}$ (Table \ref{case}) and found that the development of melt-fingers does not depend on these parameters.

     In contrast to melt-fingers, the growth of partially molten plumes depends on the Rayleigh number. At $Ra = 2.15 \times 10^4$ that is lower than Ra of the reference case by a factor of 100 (Case Ra2.15e4), only melt-fingers develop, and partially molten plumes do not grow (Figure \ref{per5_e4}b). Melt-fingers grow upward to the depth level of around 75 km by around 1.0 Gyr (Figure \ref{per5_e4}b and Table \ref{case}) and then expand laterally, to make the most part of the mantle partially molten (Figure \ref{per5_e4}b for 1.04 Gyr). After that, the partially molten region shrinks monotonously with time owing to conductive cooling from the surface boundary (Figure \ref{per5_e4}b for 2.80 Gyr). We found that partially molten plumes grow only in the cases calculated at $Ra \geq 7.15 \times 10^5 $ (see Figure \ref{pngfiguresample2}). This result reinforces that partially molten plumes are driven by their buoyancy.

   \subsection{The effect of the initial distribution of mantle composition}
\begin{figure}
\noindent\includegraphics[width=\textwidth]{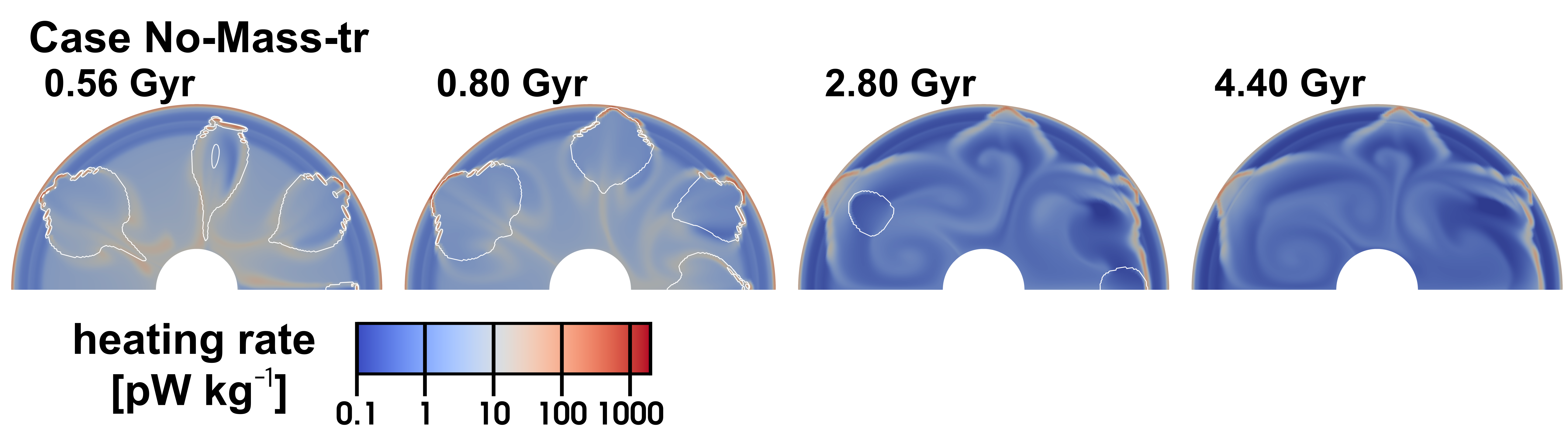}
\caption{Snapshots of the distributions of internal heating rate $q$ calculated in a case where the composition $\xi_\mathrm{b}$ is kept uniform (Case No-Mass-tr). The contour lines show the distribution of melt-content with contour level of 0 and 0.1.}
\label{no-mass}
\end{figure}
    \noindent To see how plume magmatism continues for billions of years in the reference case (Figure \ref{for_journal}), we further calculated Case No-Mass-tr where we assumed a compositionally uniform mantle with the bulk composition $\xi_{\mathrm{b}}=\xi_{\mathrm{e}}$ in the initial condition (Table \ref{case}). 
    In this case, most of partially molten plumes ascend to the uppermost mantle by around 0.8 Gyr, and plume activity declines after that (Figure \ref{no-mass}). 
    The early decline of plume activity results from the early extraction of HPEs from the deep mantle where magma is mostly generated. In contrast, HPE-extraction by early plume magmatism is not so efficient in the reference case where the deep mantle is assumed to be enriched in the dense IBC-component in the initial condition (see Section \ref{Thermal and structural evolution of the mantle}). The duration time of plume activity depends on the initial distribution of the dense IBC-component.

   \subsection{Dependence of radial expansion on model parameters} 
   \label{Radial expansion/contraction history of planets}
   
\begin{figure}
\noindent\includegraphics[width=\textwidth]{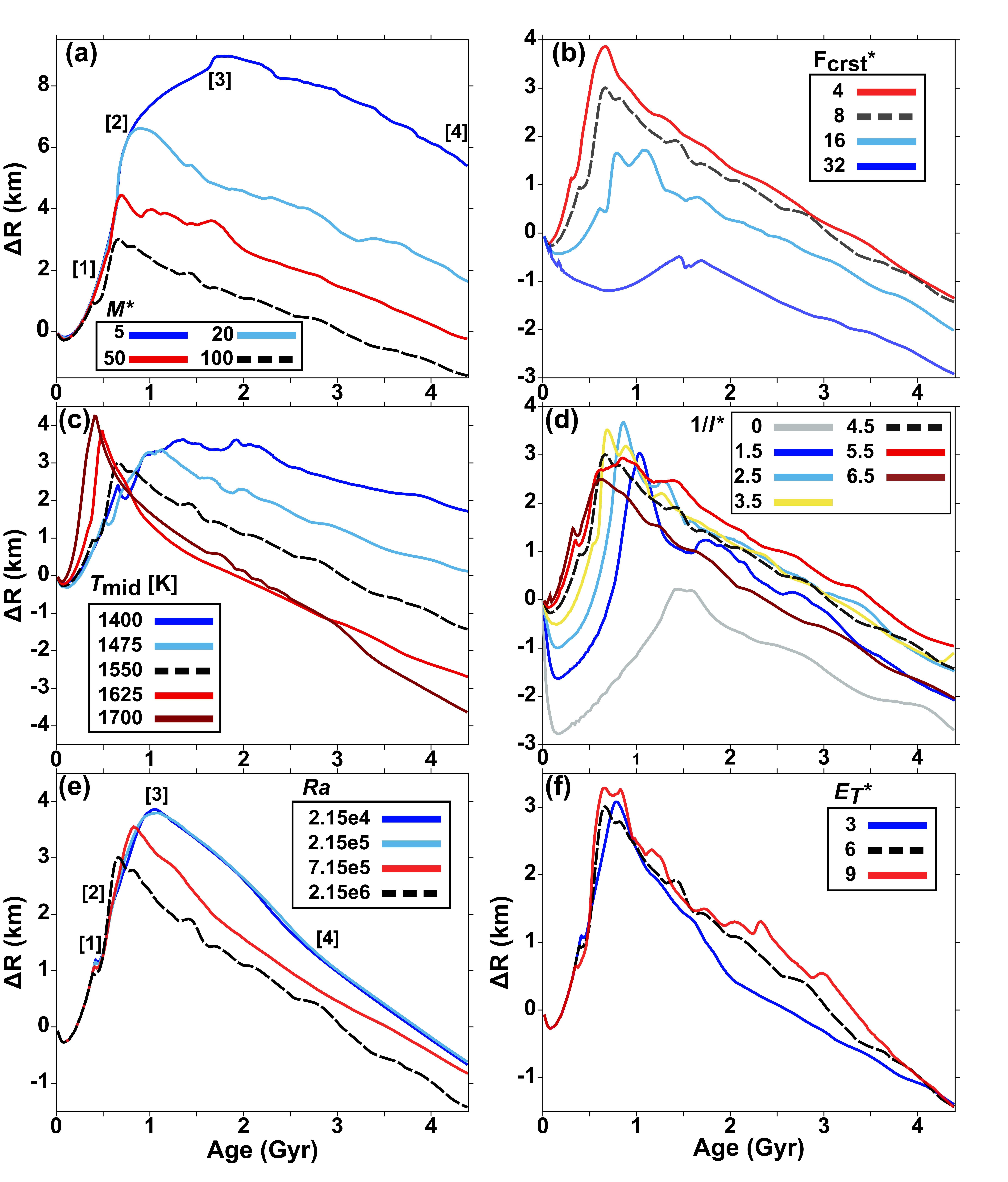}
\caption{Plots of radius change against time calculated at various values of (a) the reference permeability $M^*$, (b) the initial crustal fraction of the HPEs ${F_{\mathrm{crst}}}^*$, (c) the initial temperature in the mid mantle $T_{\mathrm{mid}}$, (d) the thickness of the overturned layer $l^*$, (e) the Rayleigh number $Ra$, and (f) the sensitivity of viscosity to temperature $E_T^*$. The dotted lines show the reference case (Case Ref) presented in Figure \ref{for_journal}-\ref{radius}. In (a) and (e), the numbers [1] to [5] correspond with those of Figure \ref{per5_e4}a and b, respectively.}
\label{all_radius}
\end{figure}

   \noindent We further carried out numerical experiments to show how the radial expansion-history depends on the model parameters: the reference permeability $M^*$; the Rayleigh number $Ra$; the sensitivity of viscosity to temperature $E_T^*$; the initial mid-mantle temperature $T_{\mathrm{mid}}$; the thickness of the overturned layer $l^*$; the initial concentration ratio of the HPEs in the crust to the mantle $F_{\mathrm{crst}}^*$ (Figure \ref{all_radius} and Table \ref{case}). 

   Among the parameters, the reference permeability $M^*$ influences the expansion-history most strongly (Figure \ref{all_radius}a); a lower $M^*$ leads to a later peak of expansion with a larger amplitude. The larger expansion is due to a more slowly migrating magma that retains HPEs in the deep mantle for a longer period, and allows more heat to build up in the mantle (see Figure \ref{per5_e4}a and Table \ref{case}). 
   The initial ratio of HPE-concentration in the crust to that in the mantle $F_{\mathrm{crst}}^*$ also substantially influences the magnitude of radial expansion (Figure \ref{all_radius}b); a higher $F_{\mathrm{crst}}^*$ reduces radial expansion because the mantle is more depleted in HPEs and less magma is generated in the deep mantle. 
    %The dependence of expansion-history on other model parameters is not so significant as that on $M^*$, $F_{\mathrm{crst}}^*$, and $l^*$.
   The initial mid-mantle temperature $T_{\mathrm{mid}}$ affects only the timing of radial expansion as shown in Figure \ref{all_radius}c. A higher $T_{\mathrm{mid}}$ leads to an earlier peak of expansion because it implies an earlier generation of partially molten regions in the deep mantle and earlier extension of the regions into the uppermost mantle. 
    The thickness of the overturned layer $l^*$ has an effect on the beginning of radial expansion (Figure \ref{all_radius}d). A higher $1/l^*$ where the deep mantle is more enriched in HPEs induces an earlier beginning of the expansion. Note that in Case l0 where HPEs and the composition $\xi_{\mathrm{b}}$ are uniform in the whole mantle ($1/l^*=0$), the amplitude of that is smaller than in cases where the overturned layer is considered (Figure \ref{all_radius}d). This is because enough magma generation does not occur to cause the early expansion in Case l0.
   Conversely, although the horizontal averages of the temperature-distributions depend on the values of $Ra$ and $E_T^*$ (Figure \ref{distribution}a), the dependences of expansion-history on the Rayleigh number $Ra$ and the sensitivity of viscosity to temperature $E_T^*$ are negligible as shown in Figure \ref{all_radius}e and f. 
   %The speed of migrating magma depends on $M$ strongly; the migrating magma transports HPEs to the uppermost mantle in the early history when $M$ is as high as 100 regardless of the values of $Ra$ and $E_T^*$.

 \section{Discussions} 
 
 \begin{figure}
\noindent\includegraphics[scale=0.5]{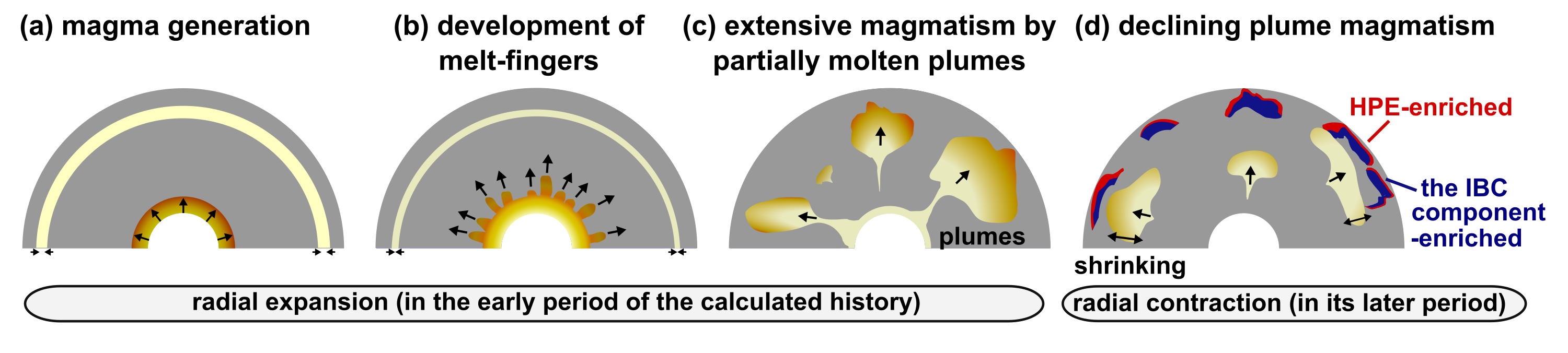}
\caption{An illustration of the thermal and structural history of the lunar mantle inferred from our numerical study. The yellow color stands for melting.}
\label{discuss}
\end{figure}

  \noindent Figure \ref{discuss} illustrates how the mantle evolves in the reference case shown in Figures \ref{for_journal}-\ref{radius}. The partially molten region in the uppermost mantle shrinks with time, whereas that in the deep mantle expands on the earliest stage of the calculated history (Figure \ref{discuss}a). 
  %The radius contracts for the first around 150 million years because of the solidification of the magma in the shallow mantle which magma assumed in the initial condition; then the planet expands radially due to the generation of the magma in the deep mantle (Figure \ref{discuss}a). 
  Melt-fingers then develop along the top of the partially molten region at depth, and the fingers extend upward (Figure \ref{discuss}b). After the growth of melt-fingers, partially molten plumes driven by melt-buoyancy rise to the uppermost mantle to cause plume magmatism (Figure \ref{discuss}c). 
  The planet expands in the stage of Figure \ref{discuss}a-c because of melting of the mantle. These partially molten plumes transport HPEs and the IBC component from the deep mantle to the uppermost mantle. The plume magmatism then declines with time after around 2 Gyr of the calculated history because of depletion of HPEs in the deep mantle (Figure \ref{discuss}d). 
 In the later period of the calculated history, the planet gradually contracts by cooling and solidification of the partially molten mantle. 
The reference case fits in with the observed history of the Moon the most among the models calculated here as we will discuss in Section \ref{Comparison with the observed features of the Moon}.

\subsection{Comparisons with earlier models}
\subsubsection{The mantle evolution caused by magmatism and convection}

  \noindent A comparison with earlier classical 1-D models of lunar thermal history in the literature shows the crucial role that heat transport by migrating magma and mantle convection plays in the reference case.
   In earlier 1-D models where only internal heating and thermal conduction are considered, the temperature in the deep mantle monotonously increases to the solidus temperature due to internal heating, while the lithosphere monotonously thickens with time owing to cooling from the surface boundary throughout the calculated history \cite<e.g.,>{Wood,solomon&T,T&solomon,s&c1976}. These models show that the deep mantle becomes extensively molten and the lithosphere becomes as thick as around 600 km at present. In our reference case where heat transport by migrating magma and mantle convection is also considered, however, the temperature is below the solidus in most part of the deep mantle, and the thickness of the lithosphere is by less than around 350 km at 4.4 Gyr (Figure \ref{distribution}a).

 \begin{figure}
\noindent\includegraphics[scale=0.7]{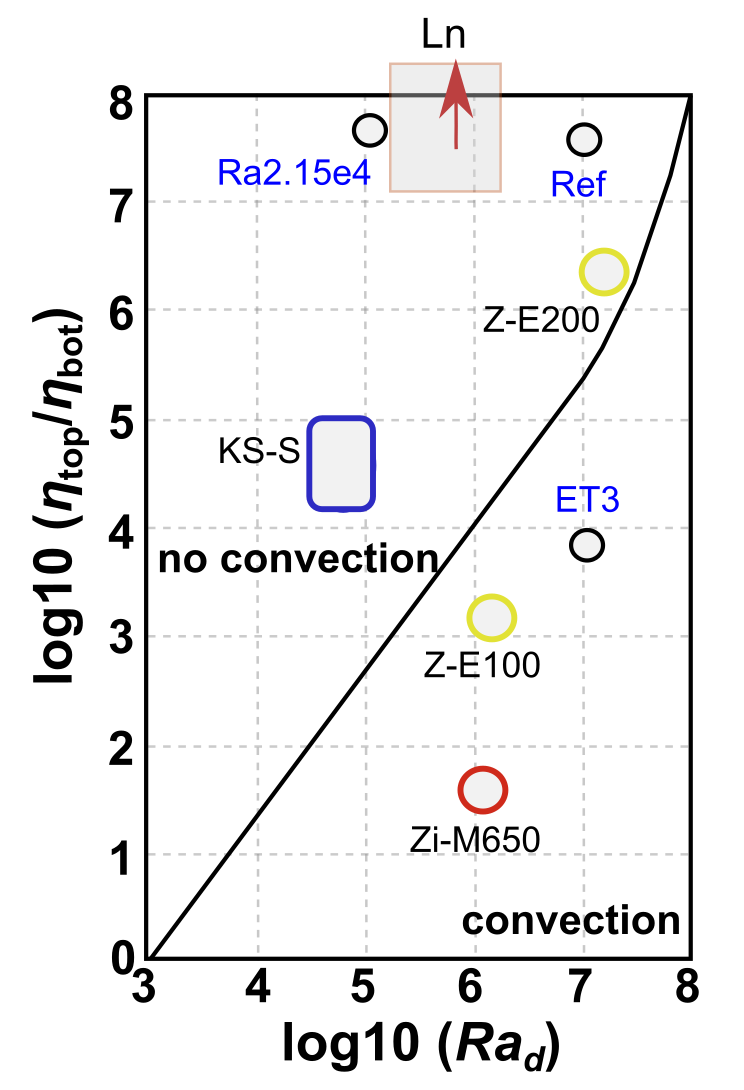}
\caption{Plots of the critical Rayleigh number for onset of thermal convection in the 3-D spherical mantle heated from the CMB, taken from Figure 8b in \citeA{yanagisawa}. This viscosity is assumed to depend on the temperature and the magnitude dependence is measured by the viscosity contrast between the surface boundary $\eta_{\mathrm{top}}$ and the CMB $\eta_{\mathrm{bot}}$. Z-E100 and Z-E200 (the yellow circles) correspond to the case H50E100v5e20 and H50E200v5e20 of \citeA{zhang2013a}; the brown arrow (Ln) by \citeA{Laneuville2013}; the red circle (Zi-M650) by the case M650 of \citeA{zie}; the blue rounded rectangle (KS-S) by \citeA{k&s,Spohn}. The black circles (Ref, Ra2.15e4, and ET3) also correspond to the Case Ref, Ra2.15e4, and ET3.}
\label{Yanagi}
\end{figure}

A comparison with earlier 2- or 3-D models of the lunar thermal history where mantle convection also is considered, on the other hand, shows the crucial roles that magma plays in heat transport in the convecting mantle. The vigor of thermally driven mantle convection is controlled by the distance from the threshold for the onset of thermal convection on the plane of the Rayleigh number $Ra_d$ and the viscosity contrast $\eta_{\mathrm{top}}/\eta_{\mathrm{bot}}$ shown in Figure \ref{Yanagi} \cite{yanagisawa}.
As shown in the figure, our reference case (Case Ref) is calculated under the condition that thermally driven mantle convection is more sluggish than that of the earlier models of lunar mantle convection Z-E100, Z-E200 \cite{zhang2013a}, and Zi-M650 \cite{zie}.
%, similar to that of Case Ra2.15e4 ($E_T^*$=3).
The lithosphere at 4.4 Gyr in our reference case is (see the black line in Figure \ref{distribution}a), however, thinner than that in these models (see e.g., Figure 3c in \citeA{zhang2013a}). 
Even in the models where thermally driven mantle convection does not occur at all (Case Ra2.15e4), the lithosphere (see the red line in Figure \ref{distribution}a) is substantially thinner than that in the earlier models KS-S \cite{k&s,Spohn} and Ln \cite{Laneuville2013}. These differences arise because, in our models, the uppermost mantle is kept hot by melt-fingers and partially molten plumes owing to the matrix-expansion around the fingertip and the melt-buoyancy, respectively. 
Note that the lithosphere of Case Ra2.15e4 is also thinner than that in a spherically symmetric 1-D model of magma migration in the non-convecting mantle \cite{u2022} because of heat transport by melt-fingers. Heat transport by melt-fingers and partially molten plumes is an essential part of our models. 

Because of the explicit implementation of a model of magma generation and migration into that of mantle convection, our models allow us to infer volcanic history directly from the calculated history of mantle melting. 
Earlier models have discussed volcanic history based on the calculated distribution of partially molten regions, especially the depth of the top of the regions and the calculated rate of magma generation in the mantle 
\cite<e.g.,>{Wood,solomon&T,PKT,Spohn,Laneuville2018,u2022}.
These models assume that the magma ascends from partially molten regions as deep as 200-800 km to the crust. However, it is not clear if magma eruption is well correlated with the depth and magma generation rate. Magma may not be able to make its way to the surface when the stress state in the lithosphere is horizontally compressive and hence when the planet is contracting \cite{Solomon&head,Solomon1986}. 
In our models, in contrast, magma rises directly to the base of the crust by melt-fingers and partially molten plumes (Figures \ref{for_journal} and \ref{distribution}), allowing us to predict volcanic history more convincingly. Several issues, however, still remain on magma migration in the crust and the uppermost mantle. It is unclear how mare basaltic magma can ascend through the crust that is not denser than the magma \cite<e.g.,>{HW1992}. The detail of formation of dikes in the uppermost mantle is also important for understanding the magma ascent \cite<e.g.,>{WH2003,WH,HW}. These issues may be important for also constructing a more refined model of thermal history of the Moon as noted by \citeA{Lourenco}: the thermal history can depend on the ratio of extrusive to intrusive volcanism, which can substantially depend on the porosity, thickness, and density of the crust \cite<e.g.,>{Solomon1975,morota2009,Taguchi,H&W2020}. In future studies, a more refined modeling of magma migration through the crust and the uppermost mantle is needed. 

In order to understand the evolution of the compositionally stratified mantle predicted from the hypotheses of the magma ocean and mantle overturn, mass transport by migrating magma and magma-driven convective flow of the mantle is indispensable.
The lunar mantle is expected to have been compositionally stratified with a layer enriched in the compositionally dense IBC-component at the base of the mantle after putative crystal fractionation in the magma ocean and subsequent mantle overturn \cite<e.g.,>{snyder,hess&P,Elkins-Tanton}. 
Some mantle convection models suggest that the basal layer eventually rises as upwelling plumes owing to its thermal buoyancy as the layer is heated by HPEs \cite<e.g.,>{zhong,stegman,zhang2017,zhang2022}. In these models, however, the excess composition of the basal layer with respect to the overlying olivine-rich mantle is less than that suggested by some recent models of lunar mantle overturn \cite<e.g.,>{yu,zhao2019}, greater than $160$ kg $\mathrm{m^{-3}}$ (see Figure 6 in \citeA{li2019}).
At this large compositional density contrast, the basal layer is convectively stable and does not ascend as plumes by thermal buoyancy alone, as inferred from earlier laboratory experiments of mantle convection (see Figure 1 in \citeA{Bars}).
    Although the compositional density contrast in our reference case ($l^*=4.5$) is around $180$ kg $\mathrm{m^{-3}}$ (Figure \ref{initial}), a large fraction of the IBC-rich materials in the layer is extracted and transported to the uppermost mantle by melt-fingers and by partially molten plumes driven by melt-buoyancy (Figure \ref{for_journal}); mass transport by migrating magma and melt-buoyancy driven convective flow of the mantle are crucial for understanding the structural evolution of the mantle.

    Our model also shows that a careful modeling of HPE transport is important to correctly understand the thermal and volcanic history of lunar mantle. The lunar volcanism has continued until 1-2 Gyr ago \cite<e.g.,>{hiesinger2000,whitten&head}, and some earlier mantle convection models conclude that the volcanism has continued for such a long period because partially molten regions remained in the cooling mantle for billions of years \cite{k&s,Spohn,zie}. In models where the uppermost mantle is locally more enriched in HPEs, the partially molten regions observed in there persist for more than 3 Gyr \cite{Laneuville2013,Laneuville2014,Laneuville2018}.
   Volcanism, however, extracts HPEs from the mantle to let the decline of subsequent volcanic activity \cite{cassen1973,cassen1979,ogawa2014,ogawa2018B}. In particular, \citeA{ogawa2018B} suggests that partially molten regions disappear within 2 Gyr since the beginning of the calculated history, too short to be a model of the lunar mare volcanism, owing to extraction of HPEs by magmatism. In our models, in contrast, magmatism continues for a much longer time despite that HPE transport by magma is considered (Figures \ref{for_journal} and \ref{distribution}) because a compositionally dense IBC-enriched layer is assumed at the base of the mantle in the initial condition. When HPE- and IBC-enriched materials in the basal layer are transported by melt-fingers and partially molten plumes, the magma often solidifies on the way to the surface and sinks again to the deep mantle (Figure \ref{for_journal}; see also Movie S1-4). The initial mantle stratification is necessary for magmatism to continue long. Indeed, in Case No-Mass-tr where a compositionally uniform mantle is assumed in the initial condition ($\xi_{\mathrm{b}}=\xi_{\mathrm{e}}$), the mantle becomes completely solid, and magmatism declines much earlier than that in the reference case (Figure \ref{no-mass}).

For a more realistic simulation of the evolution of lunar mantle, it is essential to extend the model to a 3-D spherical shell \cite<e.g.,>{Laneuville2013,zhang2017}. 2-D annular models of mantle convection tend to predict a higher average temperature in the mantle than 3-D spherical models do, especially when the core size is small, as is the case for the Moon \cite{Guerrero}. Although it is computationally challenging, modeling magma generation and migration in a 3-D spherical mantle is a promising avenue for future research.

The dichotomy is also a long-standing issue in studies of mantle dynamics in the Moon \cite<e.g.,>{Jolliff,Lawrence,cho2012}. Our models show a localized compositional and thermal heterogeneity in the uppermost mantle (Figure \ref{for_journal}), but not a global-scale dichotomy like the one that results from an upwelling mantle plume observed in the model of \citeA{zhang2013a}. Some studies suggest that the lunar dichotomy is caused by an exogenous agent, such as the South Pole-Aitken impact \cite<e.g.,>{arai,jones,zhang2022nature}. It is important to assume the initial condition in which the uppermost mantle in the nearside is more enriched in HPEs than that in the farside \cite{PKT,Laneuville2013,Laneuville2014,Laneuville2018}.

  \subsubsection{The radial expansion/contraction}
 \noindent Our reference case shows that volume change of the mantle by melting is a key for understanding the radius change of the mantle. 
 To cause the observed early expansion of the Moon, classical earlier models suggest that the initial temperature in the deep mantle was less than 1200 K \cite{s&c1976,Solomon1986,kirk&s}. This upper limit is, however, substantially lower than that expected from earlier models of mantle overturn that start from giant-impact hypotheses in the literature \cite<e.g.,>{p&s,canup,lock}. These overturn models suggest that the initial temperature in the deep mantle is approximately 1800-1900 K \cite<e.g.,>{hess&P,Elkins-Tanton,bouk,li2019}. 
  When such a high initial temperature is assumed, the early expansion occurs only nearside \cite{Laneuville2013} or is much smaller than the observed expansion \cite{zhang2013a}. In a model where the blanket effect in the crust is taken into account \cite{zie,zhang2013b}, thermal expansion of more than 1 km does occur but continues for longer than 1 Gyr, too long to account for the radial expansion of the Moon \cite<e.g.,>{hana2013}; thermal expansion of the planet continues for such a long period even in our reference case, as shown by the blue line in Figure \ref{radius}.
  In our model, however, global expansion occurs by a few kilometers within the first 0.7 Gyr of the calculated history because of melting of the mantle. The radial expansion by melting occurs earlier and is larger than the thermal expansion (Figure \ref{radius}), suggesting that mantle melting dominates the radial expansion/contraction history of the Moon.

  The radial expansion/contraction history depends on the spatial dimensionality. 
 In a 1-D spherically symmetric model where volume change by melting of the mantle is considered, the amplitude of radial expansion is around 1.2 km (see Figure 2 in \citeA{u2022}). In contrast, the amplitude in our 2-D model is around 3 km (Figure \ref{radius}), substantially larger than that in the 1-D model. To predict more quantitatively the radial expansion/contraction history of the Moon, it is necessary to develop a model in a 3-D spherical shell \cite{Laneuville2013,zhang2013a,zhang2013b}.

\subsection{Comparison with the observed features of the Moon}
\label{Comparison with the observed features of the Moon}
\subsubsection{The radial expansion/contraction}
 \noindent The planet globally expands by a few kilometers for the first several hundred million years in our reference case owing to the volume change caused by melting of the mantle (Figure \ref{radius}). The timing and amplitude of radial expansion are consistent with those of the Moon inferred from its gravity field in the literature \cite{hana2013,hana2014,sawada,liang&Andrews}. 
 After the radial expansion, the planet begins to radially contract at around 1 Gyr after the start of the calculation (Figure \ref{radius}). The timing of contraction is consistent with the beginning of compressive tectonics on the Moon \cite{Yue,Frueh}. The planet in our model contracts at a rate of approximately $-1.0 \; \mathrm{km}\;\mathrm{Gyr^{-1}}$ in the past billion years (Table \ref{case}), which is also consistent with the estimates obtained from observations of fault scarps on the Moon \cite<e.g.,>{watters,Klimczak,clark,bogert}. As a whole, the calculated history of radius change of the reference case is consistent with that of the Moon.
 %The contraction begins around 1 Gyr after the start of the calculated history, which is consistent  \cite{Solomon1986,kirk&s,zhang2013a,zhang2013b}. However, observations suggest that the fault scarps were active only within the last billion years \cite<e.g.,>{watters2015,clark,bogert}.
 %It would suggest that the faults were formed as a result of global stresses that began to accumulate more than 3 Gyr ago \cite{s&c1976,Solomon&head,binder&gunga,Frueh}.
   Among our models presented in Figure \ref{all_radius}, the radius changes calculated at $M^* \geq 50$, $T_{\mathrm{mid}} \approx 1550$ K, and ${F_{\mathrm{crst}}}^* \leq 16$ are consistent with the observed one. When the reference permeability $M^*$ is lower than 50, the amplitude of radial expansion is much larger than the estimate for the Moon because generated magma in the deep mantle stays there for a long time and is not extracted by melt-fingers (Figures \ref{per5_e4} and \ref{all_radius}a); melt-fingers are most likely to play an important role in regulating the history of radius change in the Moon.
   In the cases where the initial concentration ratio of the HPEs in the crust to the mantle ${F_{\mathrm{crst}}}^*$ is larger than 16 (the average heating rate in the mantle is less than 7.6 $\mathrm{pW \, kg^{-1}}$ in the initial condition), sufficient partially molten regions cannot develop to cause the early expansion (Figures \ref{all_radius}b).
   Besides, in order for the early expansion to occur, a substantial fraction of the mantle should be solidified in the initial state (Figures \ref{initial}a and \ref{all_radius}c).
   %7.5827 pw以上じゃないとダメ
   %
   Our models also show that the timing and amplitude of radial expansion strongly depend on the presence of overturned IBC-component enriched layer $l^*$ (Figure \ref{all_radius}d and Table \ref{case}).
    In Case l0 where the overturned layer is not assumed ($1/l^*=0$), indeed, the timing and amplitude of calculated radial expansion are later and smaller than the observed expansion.

\subsubsection{The volcanic activity}
\noindent Our reference case is also consistent with the observed history of mare volcanism of the Moon. For the first 0.4 Gyr of the calculated history, the partially molten regions in the deep mantle grow only slowly, while the partially molten region in the uppermost mantle shrinks over time (Figure \ref{for_journal}). This early stage is likely to correspond to the period during which mare volcanism on the Moon was not so active \cite<e.g.,>{hiesinger2003,whitten&head}. The growth of melt-fingers and subsequent partially molten plumes observed in Figure \ref{for_journal}b account for the lunar mare volcanism that became active after around 4 Gyr ago and peaked at 3.5-3.8 Gyr ago, and then gradually declined over a period of billions of years \cite<e.g.,>{hiesinger2000,morota2011b}; the calculated activity of partially molten plumes is indeed peaked at around 3.7 Gyr ago and then gradually declines with time.
In contrast, in a model where plumes do not appear (Case Ra2.15e4 calculated at lower Rayleigh number $Ra = 2.15 \times 10^4$), the volcanic activity develops only during the first several million years by melt-fingers (Figure \ref{per5_e4}b), suggesting that partially molten plumes play an important role in the volcanic history of the Moon.

Note that magma not enriched in HPEs is generated in the later period of the calculated history (Figure \ref{for_journal} for 1.44 Gyr). This volcanism is caused by a return flow of a floundering materials enriched in IBC components (see the arrow in Figure \ref{for_journal}d). This may account for the volcanism of HPE-depleted young basalts \cite{Che2021,li2021,su2022}.
% (Note that our models cannot suggest local volcanic activities. Recent studies reveal that mare volcanism in the PKT was still active in the later period \cite<e.g.,>{kato2017,qian2023,wang2023}. 
 %Observations from the Lunar Reconnaissance Orbiter suggest that basaltic volcanism has occurred in a region of peculiar small depression, called Ina, within the last 100 million years \cite{Braden}. 
% We discuss volcanism on the whole Moon in this study, but a local model of the Moon is important for a proper understanding of regional lunar volcanism.)

\subsubsection{The temperature profile}
\noindent The depth-profile of the horizontally averaged temperature calculated at 4.4 Gyr in our reference case shows that the lithosphere develops as a thermal boundary layer of the convective mantle (Figure \ref{distribution}a), while the temperature profile suggested for the present Moon \cite<e.g.,>{sonett,khan2006,khan2014,karato2013} is closer to a thermal diffusion profile.
The difference is not large in the shallow mantle (until the depth level of around 200 km), but the calculated temperature in the mid-mantle is considerably higher than that in the present Moon (Figure \ref{distribution}a).
%The depth-profile of the horizontally averaged temperature calculated at 4.4 Gyr in our reference case shows a feature not observed for the temperature-profile of today’s Moon (Figure \ref{distribution}a). 
%The calculated temperature in the shallow mantle (until the depth level of around 200 km) is consistent with the observed temperature \cite<e.g.,>{khan2006,karato2013,Siegler2014,Siegler2022}.
%; however, the temperature in the depth level of around 500 km is up to 200 K higher than that in the Moon \cite<e.g.,>{khan2006}. 
As a consequence, the mantle is partially molten in the mid-mantle (from around $r=$ 1100 to 1300 km), whereas seismic evidence suggests that the partially molten region occurs only at the base of the mantle \cite<e.g.,>{Latham,nakamura,weber2011,tan}. A temperature profile consistent with the observed one was not obtained at other parameter values (Figure \ref{distribution}a). This difficulty may be a consequence of the assumed 2-D polar rectangular geometry of the convecting vessel and calls for further numerical calculations in a 3-D spherical shell where the mid-mantle tends to be more strongly cooled and mantle convection is more sluggish \cite{Guerrero}.
%The model in our study has a possibility to account for radius of low-velocity zone 560 or 580 km not only tidal heating but also residue of long-lives magnetism \cite{The model in our study has a possibility to account for radius of low-velocity zone 560 or 580 km not only tidal heating but also residue of long-lives magnetism \cite{tan}. }. 

%Recent studies suggest that the lunar interior is more enriched in water than previously thought \cite<e.g.,>{saal,colaprete,lin2016,lin2017,lin2020}. We should take the effect of water on mantle rheology and its transport by migrating magma into account in numerical models \cite<e.g.,>{ogawa2020}.
%here
\section{Conclusions}

 \noindent To understand the volcanic and radial expansion/contraction history of the Moon, we developed a 2-D polar rectangular numerical model of mantle evolution illustrated in Figure \ref{discuss}. The internally heated mantle of the model evolves by the transport of heat, mass, and heat-producing elements (HPEs) by mantle convection and migrating magma that is generated by decompression melting and internal heating.
 
 Our simulations show that magma generation and migration play a crucial role in the volcanic and radial expansion/contraction history of the Moon. 
 Magma is generated in the deep mantle by internal heating and eventually ascends to the surface as melt-fingers and partially molten plumes driven by melt-buoyancy for the first several hundred million years (Figures \ref{for_journal} and \ref{discuss}). This stage is likely to correspond to the period during which mare volcanism became active after 4 Gyr ago with the peak at 3.5-3.8 Gyr ago \cite<e.g.,>{whitten&head}.
 Subsequent magma ascents by partially molten plumes decline with time but continue for billions of years after the peak because some materials that host HPEs are enriched in the ilmenite-bearing cumulates (IBC) and remain in the deep mantle by their negative buoyancy (Figure \ref{for_journal}).
 This activity accounts for the lunar mare volcanism that gradually declined after the peak \cite<e.g.,>{hiesinger2003}.
 The model which accounts for the observed mare volcanism is also consistent with the radial expansion/contraction history of the Moon, which globally expanded in its earlier history until around 3.8 Gyr ago and then contracted with time \cite<e.g.,>{hana2013,Frueh}. In our model, the planet expands by a few kilometers for the first several hundred million years and then contracts over time (Figure \ref{radius}). The planetary expansion is due to the extension of partially molten regions by melt-fingers that extract magma generated in the deep mantle (Figures \ref{per5_e4}a and \ref{all_radius}); the subsequent contraction is caused by solidification of the regions due to cooling from the surface boundary.
 The early expansion by mantle melting suggested here implies that a substantial fraction of the mantle should have been solid, and there was a layer enriched in HPEs and IBC components at the base of the mantle in the Moon at the beginning of its history (Figures \ref{no-mass}-\ref{all_radius}c, d and Table \ref{case}). 
 %Compared with the case with the Rayleigh number $Ra$ is $ 2.15 \times 10^4$ (Figure \ref{per5_e4}b), the Rayleigh number $Ra \approx 2.15 \times 10^6$ is more plausible in our models to cause the continuous activity of magma ascent.
Further refinement of our model is needed to better understand the thermal history of the Moon. A global-scale lunar dichotomy like the one observed for the Moon does not arise in our models (Figure \ref{for_journal}). Some studies propose that it results from external factors, such as the impact that formed South Pole-Aitken basin \cite<e.g.,>{zhang2022nature}. In order to construct a more realistic thermal history model of the Moon, it is necessary to extend our model to a 3-D spherical shell geometry and to introduce a lateral heterogeneity in the initial thermal and compositional condition in future work.
 %In the initial condition, we assume that the deep mantle is more enriched in HPEs and IBC components as earlier models of mantle overturn suggest (Figure \ref{initial}).

\appendix
\section{The basic equations} 

\noindent In this section, we describe the basic equations that are not described in the section ``Model description’’. 
The continuity equation is
    \begin{linenomath*}
    \begin{equation}
    \label{cont_eq}
     \nabla\cdot  \mathbf{U} = -\nabla\cdot
      \left[\phi \left( \mathbf{u}-\mathbf{U}
      \right)\right]\,.
   \end{equation}
   \end{linenomath*}
   
  The momentum equation for mantle convection is
      \begin{equation}
     -\nabla P +  \rho g + \nabla\cdot \left[ \eta \left( \nabla \mathbf{U} + ^t\nabla \mathbf{U} \right)\ \right]\ = 0\,.
   \end{equation}
where the superscript $t$ means transpose of a matrix. 
  %\noindent where $\eta_{0}$ is the reference viscosity, and $T_{ref} = 1573  \;  \mathrm{K}$ is the temperature when $\eta = 10^{21}$ Pa s, and \;$E_{T} = 1.13 \;  \mathrm{K^{-1}}$ is the sensitivity of viscosity to temperature. The momentum equation in its dimensional form is
     Migration of magma is calculated in the energy equation \cite{katz2008},
     \begin{eqnarray}
     \label{energy}
   \frac{\partial \left( \rho_{\mathrm{0}} h \right)}{\partial t} + \nabla\cdot
      \left(\rho_{\mathrm{0}} h \mathbf{U}\right)= & -\nabla\cdot
      \left[\rho_{\mathrm{0}} h_{\mathrm{l}} \phi \left( \mathbf{u}-\mathbf{U} \right)\right]  - \frac{\Delta v_{\mathrm{l}}}{v_{\mathrm{0}}} \rho_{\mathrm{0}}
      {g} \phi {u_r} + 
      \nabla\cdot
      \left( k \nabla T \right) \nonumber \\ & 
      + \rho_{\mathrm{0}} q  
      + \nabla\cdot \left[ \kappa_{\mathrm{edd}} 
       \nabla \left( 
      \rho_{\mathrm{0}} h \right) \right]
     \,,
\end{eqnarray}
where $h_{\mathrm{l}}=h (\phi =1)$. $k = \rho_\mathrm{0} C_p \kappa$ is the thermal conductivity, and $\kappa$ is the thermal diffusivity. We assumed that the matrix disintegrates and that a strong turbulent diffusion occurs with the eddy diffusivity of $\kappa_{\mathrm{edd}} = 100 \kappa$ in largely molten region with $\phi > 0.4$ \cite{Kameyama, ogawa2020}. $\kappa_{\mathrm{edd}} $ is assumed to gradually increase with increasing $\phi$ as $\phi^3$ \cite{ogawa2018,u2022}. We also assumed that the crustal thermal diffusivity is 0.48 times smaller than the mantle thermal diffusivity, taking into account the blanket effect of the crust and regolith layers \cite{zie}.

We calculate the phase diagram of the binary eutectic materials $A_\xi B_{1-\xi}$ as well as the temperature, and melt-content from the chemical potential defined for the materials. In the solid phase, the chemical potential $\mu^{solid}$ is
     \begin{equation}
     \label{potential1}
      \mu^{solid} = - C_p T/\sigma_A-T\left( S_0 +C_p \;\mathrm{ln} \:T /\sigma_A\right) + P/\rho_0 \,
   \end{equation}
   for both of the end-members $A$ and $B$ where $S_0$ is an arbitrary constant. In the liquid state where the melt behaves ideal solution in this model, the chemical potential $\mu^{liquid}$ of the end-members $i$ $(i = A, B)$ is
        \begin{equation}
           \label{potential2}
      \mu_i^{liquid} = \mu_i^{l0} + RT \; \mathrm{ln} \: \xi_i^{liquid}\,,
   \end{equation}
   and
           \begin{equation}
                \label{potential3}
      \mu_i^{l0} = \mu^{solid} + \Delta h /\sigma_i \left( 1+G-T/T_i^{l0}\right)  \,.
   \end{equation}
   
   \noindent Here, $R$ is the gas constant $8.3 \; \mathrm{J \; mol^{-1} \; K^{-1}}$; $\sigma_i$ the molar mass of each end-members $(A=140.69 \;\mathrm{g \:mol^{-1}}; \;B=151.71 \;\mathrm{g \:mol^{-1}})$; $T_i^{l0}$ the dry solidus of the end member $i$ at zero-pressure. 
   %which value are solved the solidus temperature $T_{\mathrm{solidus}}$ and the composition of eutectic point of the magma $\xi_{\mathrm{e}}=0.1$ $\left( A_{0.1}B_{0.9}\right)$ as
   % \begin{equation}
    % T_{\mathrm{solidus}}= \frac{\Delta h /\sigma_A \left( 1+G\right)}{R \; \mathrm{ln} \: \left[ \exp{\left(\frac{\Delta h}{\sigma_A R T_A^{l0}}  \right)} + \exp{\left( \frac{\Delta h}{\sigma_B R T_B^{l0}} \right)}\right]}
    %   \,,
   %\end{equation}
    %   \begin{equation}
    % \xi_{\mathrm{e}}= \frac{\exp{\left(\frac{\Delta h}{\sigma_A R T_A^{l0}}  \right)}}{\exp{\left(\frac{\Delta h}{\sigma_A R T_A^{l0}}  \right)} + \exp{\left(\frac{\Delta h}{\sigma_B R T_B^{l0}}  \right)}}
   %    \,.
   %\end{equation}

The core is regarded as a heat bath of a uniform temperature $T_{\mathrm{c}}$ that changes with time as
     \begin{equation}
      c_{p\mathrm{c}} \rho_{\mathrm{core}} V_{\mathrm{core}} \frac{d T_{\mathrm{c}}}{dt} = - Sf \,,
   \end{equation}
where $c_{p\mathrm{c}} =675 $ J $\mathrm{K^{-1}}$ $\mathrm{kg^{-1}}$ \cite{zie} is the specific heat of the core, $\rho_{\mathrm{core}} = 6200$ kg $\mathrm{m^{-3}}$ \cite{kronrod2022} the core density, $V_{\mathrm{core}}$ the volume of the core, and $S$ the surface area of the CMB. The heat flux at the CMB $f$ is calculated from
     \begin{equation}
      f = - \frac{1}{\pi}  \int_{0}^{\pi} \left( k  \frac{\partial T}{\partial r}  \right) _{r=r_{\mathrm{c}}} d\theta \,.
   \end{equation}
In this study, we neglect the internal heating in the core. 
%The volume of the core is obtained by the ratio of the mantle to the core heat capacity of the Moon as
%\begin{equation}
%      \frac{c_{p\mathrm{c}} \rho_{\mathrm{core}} V_{\mathrm{core}}}{ c_{p} \rho_0 V_{\mathrm{mantle}}}  = 0.01 \,,
%   \end{equation}
%   where $V_{\mathrm{mantle}}$ is the volume of the mantle.
 %1026
 
The internal heating rate $q$ changes with time as
    \begin{equation}
      q = q_\mathrm{tr}^* q_0 \exp{\left( - \frac{t}{\tau}\right)}  \,,
   \end{equation}
where $q_0 = 14.70 $ $\mathrm{pW \, kg^{-1}}$ is the avearage initial heating rate at 4.4 Gyr ago estimated from the total amount of HPEs in the current Moon (see Table 2 in \citeA{u2022}); $\tau$ the decay time of HPEs. We approximate this value as $\tau = 1.5$ Gyr, an average of the decay times of $^{235}\mathrm{U}$ and $^{40}\mathrm{K}$ \cite{Kameyama,ogawa2020}. The non-dimensional value $q_\mathrm{tr}^*$ changs with migrating magma as
     \begin{equation}
   \frac{\partial q_\mathrm{tr}^*}{\partial t} + \nabla\cdot
      \left(q_\mathrm{tr}^* \mathbf{U}\right)= -\nabla\cdot
      \left[ q_{\mathrm{l}}^* \phi \left( \mathbf{u}-\mathbf{U} \right)\right]   + \nabla\cdot \left( \kappa_{\mathrm{edd}} 
       \nabla q_\mathrm{tr}^* \right)  \,.
\end{equation}
 Here, $q_{\mathrm{l}}^*$ is the internal heating rate in the melt as
    \begin{equation}
     \label{coeffiecient}
      q_{\mathrm{l}}^* = \frac{D q_\mathrm{tr}^*}{\left( D-1\right) \phi +1} \,,
   \end{equation}
   and, $D=100$ is the partition coefficient of HPEs between solid-phase and melt-phase.

The bulk composition $\xi_{\mathrm{b}}$ also changes with time owing to the mass transport by melt and matrix as
     \begin{eqnarray}
   \frac{\partial \xi_{\mathrm{b}}}{\partial t} + \nabla\cdot
      \left(\xi_{\mathrm{b}} \mathbf{U}\right)= & -\nabla\cdot
      \left[\xi_{\mathrm{l}} \phi \left( \mathbf{u}-\mathbf{U} \right)\right]   + \nabla\cdot \left[ \kappa_{\mathrm{edd}} 
       \nabla \left( 
      \xi_{\mathrm{b}} \right) \right] 
     \,.
\end{eqnarray}

%These equations are converted into their non-dimensional forms using the length scale $L$, the temperature scale $\mathrm{\Delta h}/C_p$, and other scaling units listed in Table \ref{non-d}.
   
%   \begin{table}
% \caption{Scaling and standard parameters}
 %\label{non-d}
 %\centering
% \begin{tabular}{l l l}
% \hline
%  Symbol  & Meaning  & Value \\
% \hline
%   $L$  & Thickness of the mantle  & $1350$ km \\
%   ${L^2}/{\kappa}$  & Heat diffusion time  & $96.32 \times 10^{9}$ year   \\
%   ${\kappa}/{L}$  & Heat diffusion velocity  & $1.40 \times 10^{-3}$ cm/year   \\
%   $\Delta h / C_p$  & Thermal change by latent heat  & $529.84$ K   \\
%   $\rho C_p \kappa \frac{\Delta h / C_p}{L}$  & Heat flow by thermal conduction  & $96.36 \times 10^{-2}$ mW $\mathrm{m}^{-2}$   \\
%   $\Delta h\frac{\kappa}{L^2}$  & Heating rate corresponding to latent heat  & $2.16 \times 10^{-1}$ pW $\mathrm{kg}^{-1}$   \\
% \hline
% \end{tabular}
% \end{table}
 
 The basic equations are converted into their non-dimensional forms using the length scale $L = r_\mathrm{p} - r_\mathrm{c}$, the temperature scale $ \Delta h / C_p$, and times scale ${L^2}/{\kappa}$. The momentum equation in its non-dimensional form is
      \begin{equation}
       \label{Ra_num}
     \nabla P^* + Ra \rho^* \mathbf{e_r} + \nabla\cdot \left[ \eta^* \left( \nabla \mathbf{U^*} + ^t\nabla \mathbf{U^*} \right)\ \right]\ = 0\,.
   \end{equation}
 where
     \begin{equation}
    \eta^* = \exp{\left[ E_{T}^* \left( T_{ref}^* - T^* \right)\right]} \,.
  \end{equation}
%Here, the asterisk stands for non-dimensional quantities ($E_{T}^* = 6$), we assume the permeability is gradually smaller with the depth as $\chi^* = \tanh \left( r^* /{\omega^*} \right) $; the thickness of the low permeable region $\omega^*$ is a free parameter in this study.
The non-dimensional relative velocity $\mathbf{u^*}-\mathbf{U^*}$ is written as
    \begin{equation}
     \label{M_num}
      \mathbf{u^*}-\mathbf{U^*} =
       -M^* g^*\frac{\phi^2}{\phi_0^3} \frac{\Delta v_{\mathrm{l}}}{v_{\mathrm{0}}} \mathbf{e_r} 
       \,.
   \end{equation}
The non-dimensional energy equation is written as
     \begin{eqnarray}
  \frac{\partial h^*}{\partial t} + \nabla\cdot
      \left(h^* \mathbf{U}\right)= & -\nabla\cdot
      \left[ h_{\mathrm{l}}^* \phi \left( \mathbf{u^*}-\mathbf{U^*} \right)\right]  - N^* g^* \frac{\Delta v_{\mathrm{l}}}{v_{\mathrm{0}}} 
     \phi {u_r^*} + 
      \nabla^2 T^* \nonumber \\ & 
      + q^*  
      + \nabla \cdot \left[ \kappa_{\mathrm{edd}}^* 
       \nabla \left( 
      h^* \right) \right]
     \,,
\end{eqnarray}
where $h^* = T^* + \phi \left( 1+G \right)$, and $N^* \equiv {{g_{\mathrm{sur}}} L}/{\Delta h}$.

\section*{Supporting materials}
This supporting information contains snapshots of the distributions of melt-content $\phi$ calculated in Case M20 and M50 (Figure \ref{pngfiguresample1}), Ra2.15e5 and Ra7.15e5 (Figure \ref{pngfiguresample2}), and animations of Case Ref presented in Figures 2-5 of the main article (Movie S1-4). See Table 3 in the main article for case numbers.

\setcounter{figure}{0}
\renewcommand{\figurename}{Fig.}
\renewcommand{\thefigure}{S\arabic{figure}}

 \begin{figure}
 \noindent\includegraphics[scale=0.8]{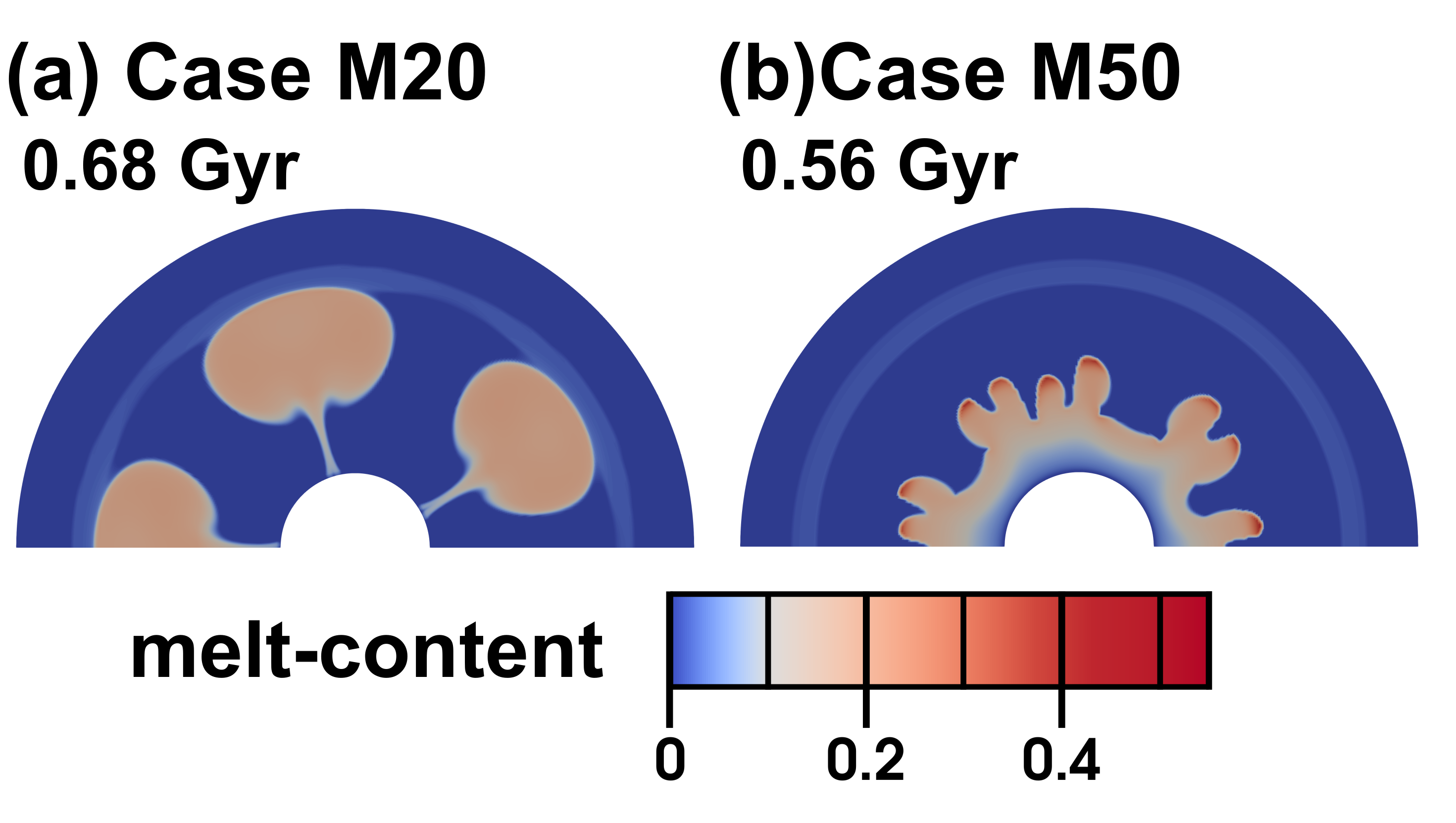}
\caption{Snapshots of the distributions of melt-content $\phi$ calculated in Case (a) M20 and (b) M50. In (b), melt-fingers develop along the top of the partially molten region at a depth.}
\label{pngfiguresample1}
\end{figure}
\begin{figure}
\noindent\includegraphics[scale=0.8]{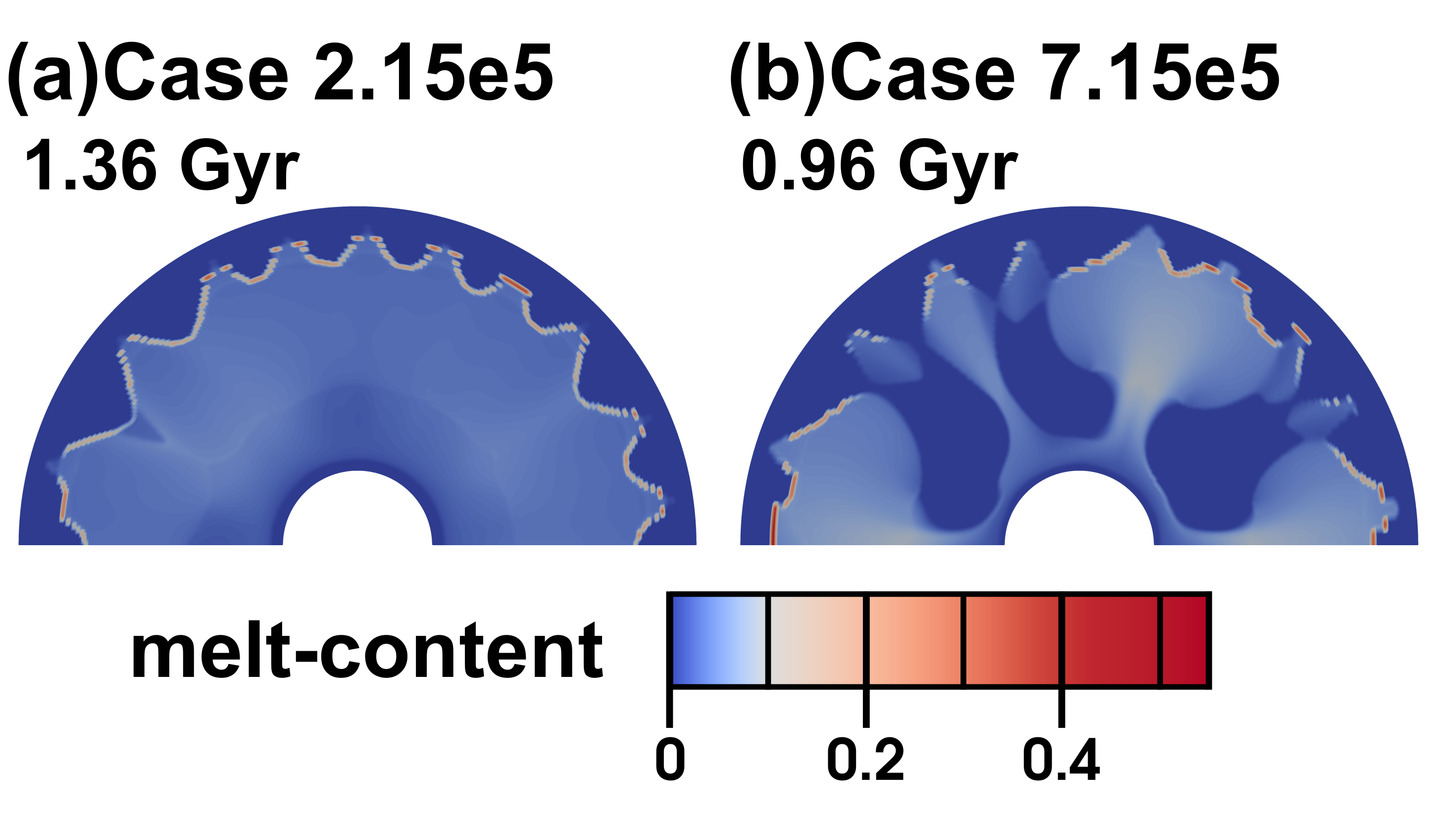}
\caption{Snapshots of the distributions of melt-content $\phi$ calculated in Case (a) Ra2.15e5 and (b) Ra7.15e5. In (b), partially molten plumes develop after the fingertip of melt-fingers reaches the uppermost mantle.}
\label{pngfiguresample2}
\end{figure}

The official version of the animations and dataset will be made available after its acceptance by the JGR: Planets. However, the unofficial version of the animations can be accessed through the following links. 

\noindent\textbf{Movie S1.} %Type or paste caption here.
%upload your movie(s) to AGU's journal submission site and select, "Supporting Information %(SI)" as the file type. Following naming convention: ms01.
\, An animation of the distributions of temperature $T$ calculated in the reference model (Case Ref). \verb|https://youtu.be/uTTyN0IANb4|
\\
\vskip.5\baselineskip
\noindent\textbf{Movie S2.} \,
An animation of the distributions of melt-content $\phi$ calculated in the reference model (Case Ref). \verb|https://youtu.be/FHcj8tZsF3Y|
\\
\vskip.5\baselineskip
\noindent\textbf{Movie S3.} \,
An animation of the distributions of internal heating rate $q$ calculated in the reference model (Case Ref). \verb|https://youtu.be/nfj9YgnUY8U| 
\\
\vskip.5\baselineskip
\noindent\textbf{Movie S4.} \,
An animation of the distributions of composition $\xi_{\mathrm{b}}$ calculated in the reference model (Case Ref). The blue color stands for the IBC component, while the red color for the olivine-rich end-member.  \verb|https://youtu.be/jRjIQ9Y_se4| 

\bibliography{agusample.bib}

\acknowledgments
%The manuscript has been greatly improved by the comments from the editor and the anonymous reviewers.
The authors would like to thank T. Yanagisawa and T. Miyagoshi at JAMSTEC for their constructive comments. This work was supported by JST SPRING, Grant Number JPMJSP2108 of Japan. Movie S1-4 and Figures \ref{for_journal}, \ref{finger_and_plume}-\ref{no-mass}, and \ref{pngfiguresample1}-\ref{pngfiguresample2} were drawn with the ParaView by Sandia National Laboratory, Kitware Inc., and Los Alamos National Laboratory.

%% ------------------------------------------------------------------------ %%
%% References and Citations

%%%%%%%%%%%%%%%%%%%%%%%%%%%%%%%%%%%%%%%%%%%%%%%
%
% \bibliography{<name of your .bib file>} don't specify the file extension
%
% don't specify bibliographystyle

% In the References section, cite the data/software described in the Availability Statement (this includes primary and processed data used for your research). For details on data/software citation as well as examples, see the Data & Software Citation section of the Data & Software for Authors guidance
% https://www.agu.org/Publish-with-AGU/Publish/Author-Resources/Data-and-Software-for-Authors#citation

%%%%%%%%%%%%%%%%%%%%%%%%%%%%%%%%%%%%%%%%%%%%%%%

%\bibliography{enter your bibtex bibliography filename here}

%Reference citation instructions and examples:
%
% Please use ONLY \cite and \citeA for reference citations.
% \cite for parenthetical references
% ...as shown in recent studies (Simpson et al., 2019)
% \citeA for in-text citations
% ...Simpson et al. (2019) have shown...
%
%
%...as shown by \citeA{jskilby}.
%...as shown by \citeA{lewin76}, \citeA{carson86}, \citeA{bartoldy02}, and \citeA{rinaldi03}.
%...has been shown \cite{jskilbye}.
%...has been shown \cite{lewin76,carson86,bartoldy02,rinaldi03}.
%... \cite <i.e.>[]{lewin76,carson86,bartoldy02,rinaldi03}.
%...has been shown by \cite <e.g.,>[and others]{lewin76}.
%
% apacite uses < > for prenotes and [ ] for postnotes
% DO NOT use other cite commands (e.g., \citet, \citep, \citeyear, \citealp, etc.).
% \nocite is okay to use to add references from your Supporting Information
%

\end{document}